\documentclass[review]{elsarticle}

\usepackage{lineno,hyperref,epstopdf}
\usepackage{graphics}

\begin{document}

\begin{frontmatter}

\title{Quantum neural computation of entanglement is robust to noise and decoherence}

\author[myaddress1]{E.C. Behrman\corref{mycorrespondingauthor}}
\cortext[mycorrespondingauthor]{Corresponding author}
\ead{behrman@math.wichita.edu}
\author[myaddress1]{N.H. Nguyen}
\author[myaddress2]{J.E. Steck}
\author[myaddress2]{M. McCann}
\address[myaddress1]{Department of Mathematics, Statistics, and Physics, Wichita State University,
Wichita, KS 67260-0033 U.S.A.}
\address[myaddress2]{Department of Aerospace Engineering, Wichita State University,
Wichita, KS 67260-0044 U.S.A.}

\begin{abstract}
In previous work, we have proposed an entanglement indicator for a general multiqubit state, which can be ``learned'' by a quantum system, acting as a neural network. The indicator can be used for a pure or a mixed state, and the system need not be ``close'' to any particular state; moreover, as the size of the system grows, the amount of additional training necessary diminishes. Here, we show that the indicator is stable to noise and decoherence.
\end{abstract}


\end{frontmatter}

\section{Introduction\label{intro}}       

The use and manipulation of entanglement is central to the exploitation of quantum computation (see, e.g., [1-8]). The quantum system obviously ``knows'' what its own entanglement is, though extraction of that information is not obvious; thus, we use dynamic learning methods\cite{lecun,werbos} to map this information onto a single experimental measurement which is our entanglement indicator\cite{behrmanqic}. Our method does not require prior state reconstruction or lengthy optimization\cite{vedral, tamaryan}, nor must the system be ``close'' to a given entangled state \cite{filip}.  An entanglement witness emerges from the learning process. We use knowledge of the smaller two-qubit system as a means of ``bootstrapping'' to larger systems \cite{efron}. As the size of the system grows the amount of additional training necessary diminishes\cite{nabic,behrmanmulti}, unlike other methods, e.g., which require knowledge or reconstruction of the density matrix \cite{bennett2, wootters, ghz}; thus, our method potentially may be of general applicability even to large-scale quantum computers, once they are built.

In any experimental implementation, though, we need also to consider that no setup is perfect: there will always be some uncertainty due to extraneous effects.  In quantum systems there is also the problem of decoherence. Classically learning systems such as neural networks have proven fault tolerant and robust to noise; they are also famously used for noise reduction in signals\cite{wasserman}. A machine learning  approach would seem to be an excellent one for issues like noise, decoherence, or missing or damaged data. Here, we show that this is in fact the case, using as a test bed our entanglement indicator on the simple two-qubit system.

\section{Dynamic learning of an entanglement indicator\label{qnn}}	        

In previous work, we showed we could successfully train a quantum system to estimate its own degree of entanglement, by mapping a measureable output at the final time, to give an indicator of the entanglement of the prepared, initial state. Briefly the method was as follows; for full details the reader is referred to\cite{behrmanqic,nabic,behrmanmulti,singapore,behrmanieee}

We begin with the Schr\"{o}dinger equation for the time evolution of the density matrix $\rho$\cite{peres}:
\begin{equation}
\frac{d \rho}{dt} = \frac{1}{i \hbar}[H, \rho] 
\label{schr}
\end{equation}
where  $H$ is the Hamiltonian. We consider an N-qubit system whose Hamiltonian is:
\begin{equation}
H   = \sum_{\alpha=1}^{N}  K_{\alpha} \sigma_{x\alpha} + \varepsilon_{\alpha} \sigma_{z\alpha} + \sum_{\alpha\neq\beta=1}^{N}\zeta_{\alpha\beta} \sigma_{z\alpha} \sigma_{z\beta}
\label{ham}
\end{equation}
where $\{ \sigma \}$ are the Pauli operators corresponding to each of the qubits, $\{K \}$ are the tunneling amplitudes, $\{ \varepsilon \}$  are the biases, and $\{ \zeta \}$, the qubit-qubit couplings. We choose the usual ``charge basis '', in which each qubit's state is given as 0 or 1; for a system of N qubits there are $2^{N}$ states, each labelled by a bit string each of whose numbers corresponds to the state of each qubit, in order. The amplitude for each qubit to tunnel to its opposing state (i.e., switch between the 0 and 1 states) is its $K$ value; each qubit has an external bias represented by its $\varepsilon$ value; and each qubit is coupled to each of the other qubits, with a strength represented by the appropriate $\zeta$ value. Note that, for example, the operator $\sigma_{x1} = \sigma_{x}\otimes I \: ... \: \otimes I$, where there are (N-1) outer products, acts nontrivially only on qubit 1. 

The parameter functions $\{ K(t),\varepsilon(t),\zeta(t) \}$ direct the time evolution of the system in the sense that, if one or more of them is changed, the way a given initial state will evolve in time will also change, because of Eqs.~\ref{schr}-\ref{ham}. This is the basis for using our quantum system as a neural network. The role of the input vector is played by the initial density matrix $\rho(0)$, the role of the output by (some function of) the density matrix at the final time, $\rho(t_{f})$, and the role of the ``weights'' of the network  by the parameter functions of the Hamiltonian, $\{ K,\varepsilon,\zeta  \}$, all of which can be adjusted experimentally\cite{yamamoto}. By adjusting these parameters using a machine learning algorithm we can train the system to evolve in time from an input state to a set of particular final states at the final time $t_{f}$. Because the time evolution is quantum mechanical (and, we assume, coherent), a quantum mechanical function, like an entanglement witness of the initial state, can be mapped to an observable of the system's final state, a measurement made at the final time $t_{f}$. Complete details, including a derivation of the quantum dynamic learning paradigm using quantum  backpropagation\cite{lecun} in time\cite{werbos}, are given in \cite{behrmanqic}. We call this quantum system a ``quantum neural network'' (QNN).

\begin{figure}
\includegraphics[height=2in]{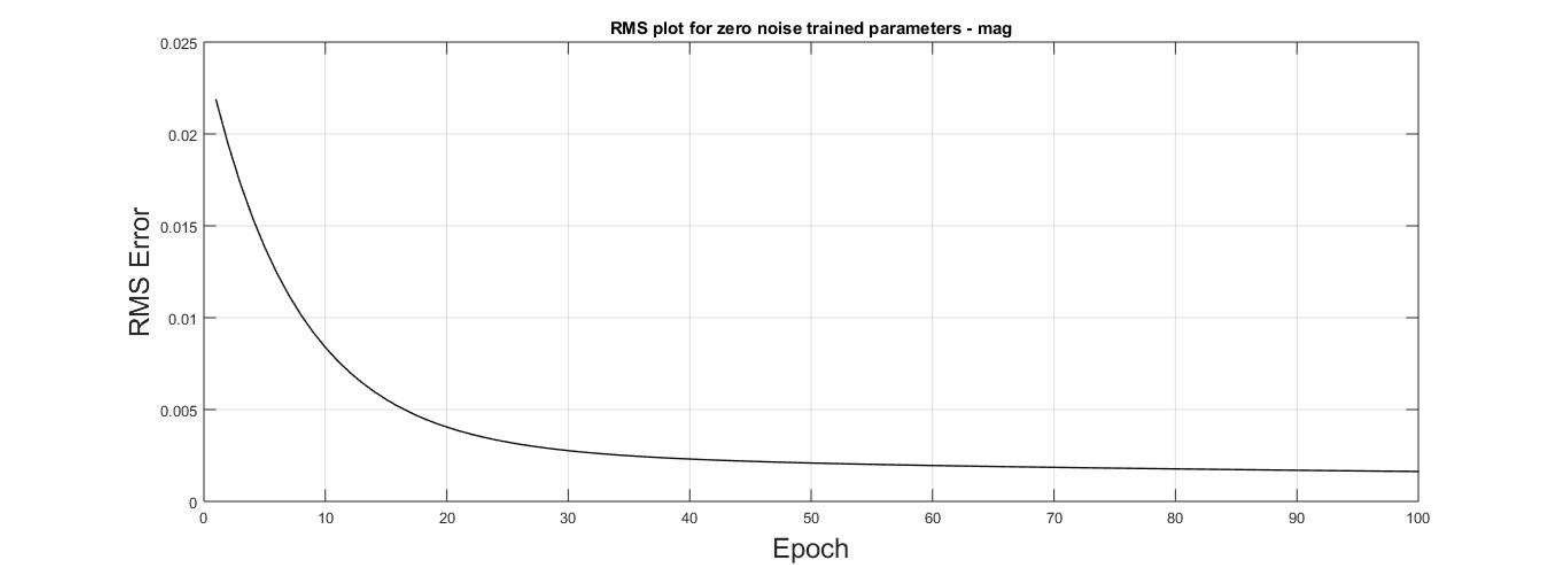} 
\caption{Total root mean squared error for the training set as a function of epoch (pass through the training set), for the 2-qubit system, with zero noise. Asymptotic error is $1.6 \times 10^{-3}$. For comparison: with piecewise constant functions a similar level of error required 2000 epochs.}
\label{trzero}
\end{figure}

\begin{figure}
\includegraphics[height=2in]{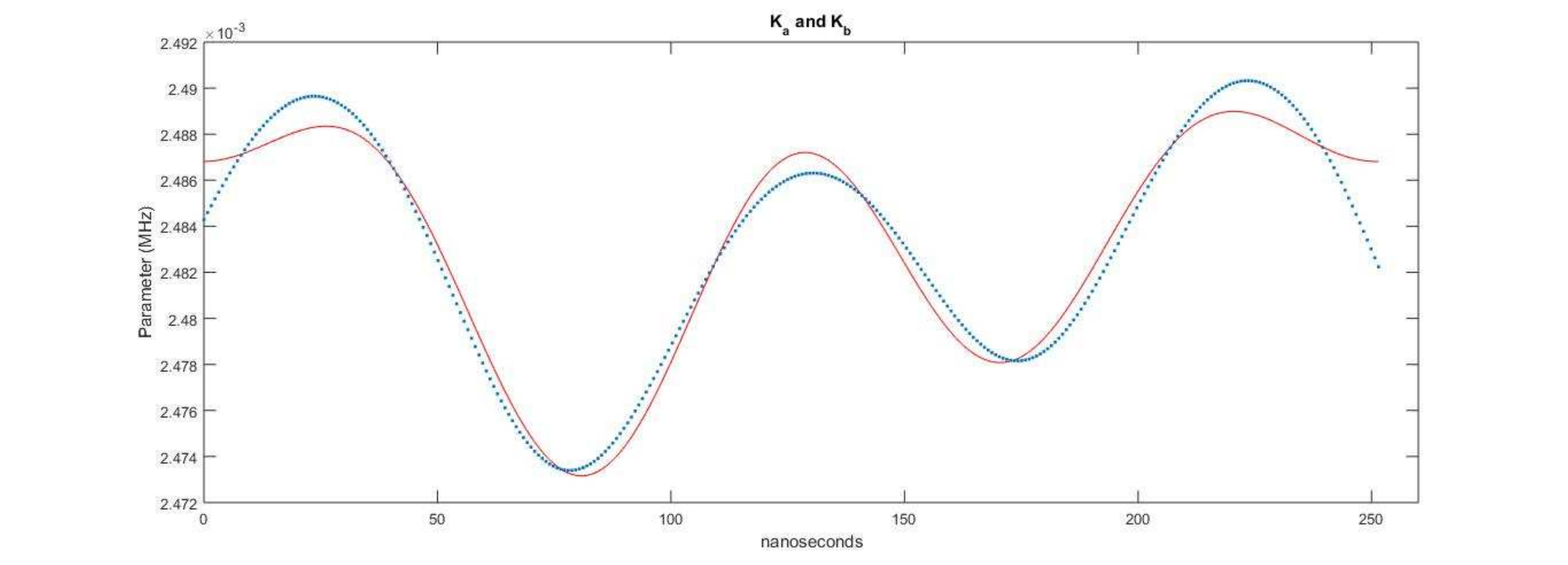} 
\caption{Parameter function $K_{A}=K_{B}$ as a function of time (data points), as trained at zero noise for the entanglement indicator, and plotted with the Fourier fit (solid line).}
\label{K0}
\end{figure}

\begin{figure}
\includegraphics[height=2in]{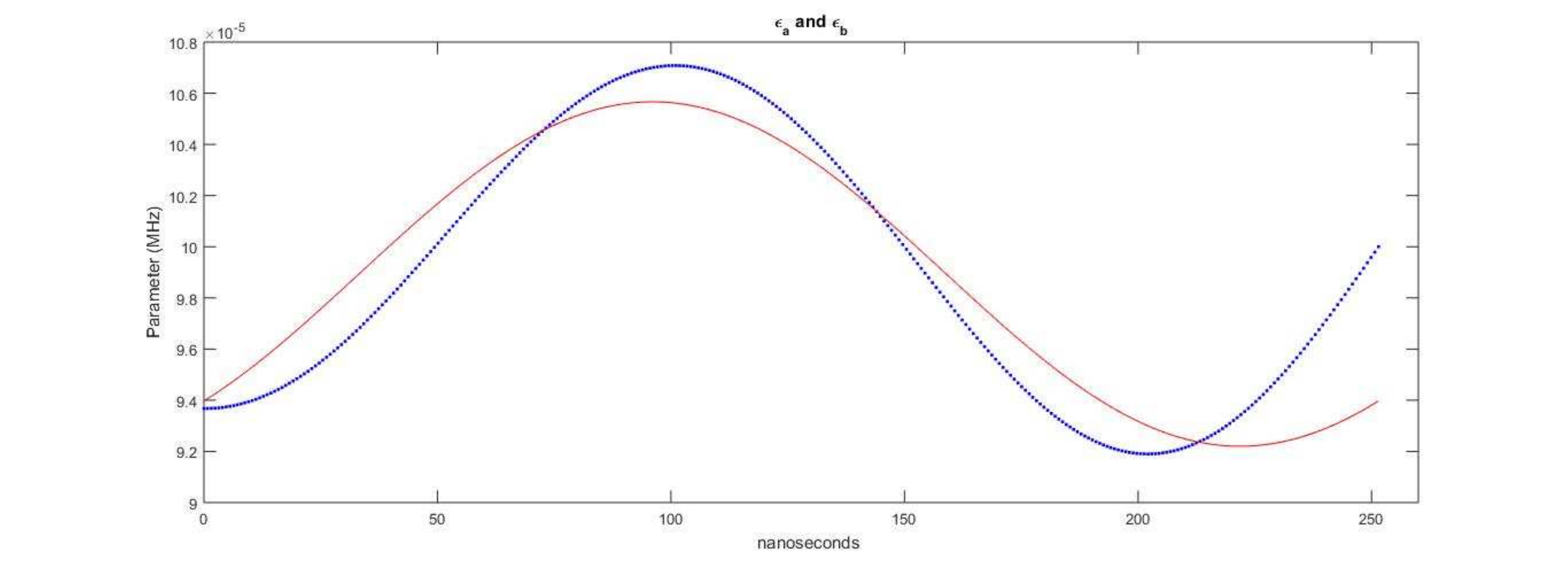} 
\caption{Parameter function $\epsilon_{A}=\epsilon_{B}$ as a function of time (data points), as trained at zero noise for the entanglement indicator, and plotted with the Fourier fit (solid line).}
\label{ep0}
\end{figure}

\begin{figure}
\includegraphics[height=2in]{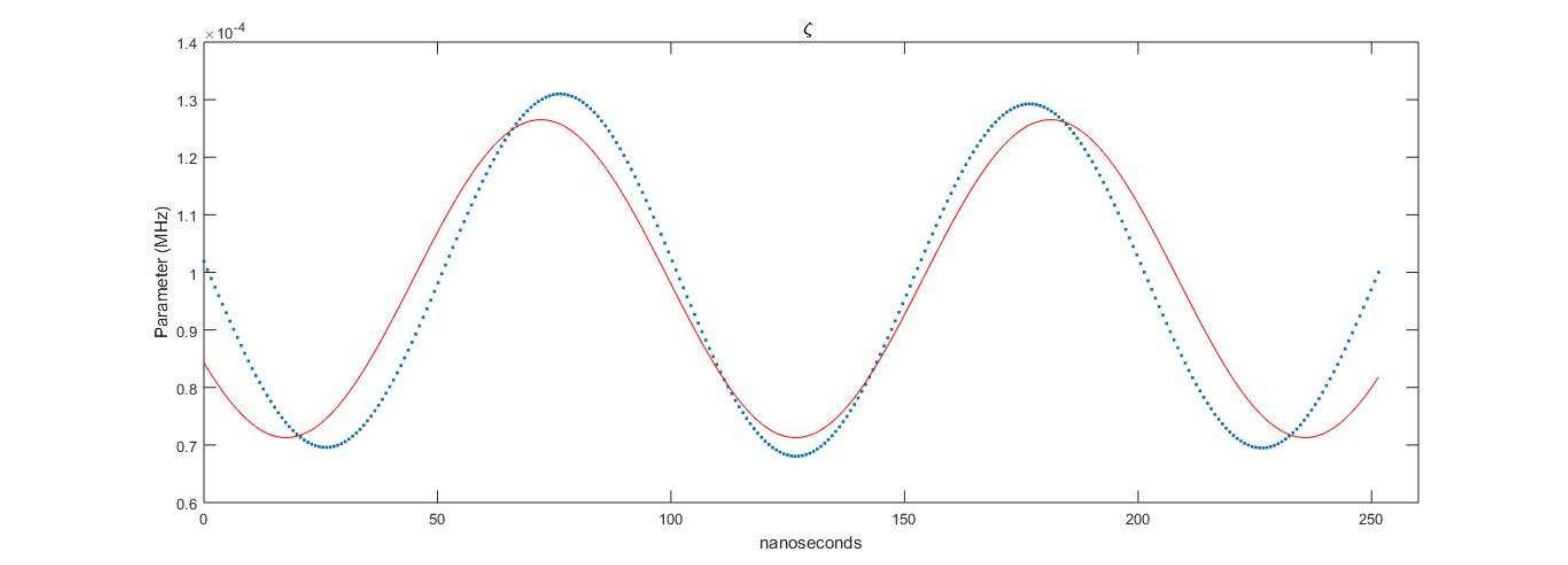} 
\caption{Parameter function $\zeta$ as a function of time (data points), as trained at zero noise for the entanglement indicator, and plotted with the Fourier fit (solid line).}
\label{z0}
\end{figure}

We  found \cite{behrmanqic, behrmanieee}  a set of parameter functions that successfully map the input (initial) state of a two-qubit system to a good approximation of the entanglement of formation of that initial state, using as the output the qubit-qubit correlation function at the final time, $\langle \sigma_{z1}(t_{f})\sigma_{z2}(t_{f})\rangle^{2}$. This set of parameter functions was relatively easily generalized to three- four- and five-qubit systems\cite{behrmanmulti}. Here, we will consider the effect of noise on only the simplest case, of two qubits ($N=2$), and for ease of notation we will call the two qubits $A$ and $B$. With continuum\cite{singapore,behrmanieee} rather than piecewise constant parameter functions\cite{behrmanqic,behrmanmulti} training is more rapid and complete; see Figure~\ref{trzero}. The parameter functions are found by training with a set of just four initial quantum states (``inputs''), as shown in Table~\ref{enttraining}: a fully entangled state (``Bell''), a ``Flat'' state (equal amounts of all basis states), a product state``C''  whose initial $(t=0)$ correlation function $\langle \sigma_{zA}\sigma_{zB}\rangle^{2}$ is nonzero, and a partially entangled state ``P''. The parameter functions are shown in Figures~\ref{K0}~\ref{ep0}~\ref{z0}. Because of the symmetry in the Hamiltonian and in the training set, $K_{A}=K_{B}=K$, and $\epsilon_{A}=\epsilon_{B}=\epsilon$. Each is a relatively simple function, well parametrized by only one frequency ($\epsilon, \zeta$) or two ($K$), as shown in Table~\ref{entparamfit} and plotted in Figures ~ref{K0},\ref{ep0},\ref{z0}. The $(input, output)$ pairs are: 

\begin{eqnarray}
input = \rho(0)= |\Psi(0)\rangle \langle \Psi(0)| \\ \nonumber
output = \langle \sigma_{zA}(t_{f}) \sigma_{zB}(t_{f}) \rangle ^{2} \rightarrow target 
\end{eqnarray} 

\noindent with prepared input states at zero time, and corresponding targets, given in Table \ref{enttraining}. This table also shows the QNN indicated entanglement values after training has finished and, for comparison, the entanglement of formation, calculated using the analytic formula \cite{wootters} for comparison. Entanglement of formation is not the only measure of entanglement, of course, but any one that we chose would have qualitatively similar behavior, which we would like our entanglement indicator to imitate. That is, we seek here not exact \underline{agreement} with $E_{F}$ (in which case we would train the state $\frac{1}{\sqrt{3}}(|00\rangle + |01\rangle + |10\rangle)$ to a target value of 0.55), but a robust and internally  self consistent measure, which we would hope would track well with an analytic measure like $E_{F}$. The QNN indicator systematically underestimates $E_{F}$ for partially entangled pure states; this is because we found through simulation that the net naturally trained to the target value of 0.44. See \cite{behrmanqic} for details. 

\begin{table}[!t]
\caption{Training data for QNN entanglement witness.}
\label{enttraining}
\begin{tabular}{l l l l}\\
\hline
Input state $|\Psi(0)\rangle$  & Target  & Trained & $E_{F}$ \\
\hline
Bell = $\frac{1}{\sqrt{2}}(|00\rangle + |11\rangle) = |\Phi_{+}\rangle$ & 1.0 & 0.998 & 1.0 \\
Flat = $\frac{1}{2}(|00\rangle + |01\rangle + |10\rangle + |11\rangle)$    & 0.0  &  $1.2 \times 10^{-5}$ & 0.0 \\
C = $\frac{1}{\sqrt{1.25}}(0.5|10\rangle + |11\rangle) $ & 0.0  &  $1.8 \times 10^{-4}$  & 0.0 \\
P = $\frac{1}{\sqrt{3}}(|00\rangle + |01\rangle + |10\rangle)$   & 0.44  & 0.44 & 0.55 \\
\hline
RMS Error   &   { }  & $1.6 \times 10^{-3}$ & { }  \\
Epochs  &  { }   &  100  &  {} \\
\hline \\
\end{tabular}
\end{table}


\begin{table}[!t]
\caption{Curvefit coefficients for parameter functions $K$, $\epsilon$, $\zeta$, for QNN entanglement witness.}
\label{entparamfit}
%\hline
$f(t) =  a_{0} + a_{1}\cos(\omega t) + b_{1}\sin(\omega t) + a_{2}\cos(2\omega t) +  b_{2}\sin(2\omega t) $ \\
%\hline
\begin{tabular}{l l l l }\\
Coefficient  & K & $\epsilon$ & $\zeta$ \\
\hline
 $a_{0}$  &   $  0.00248 $ & $9.89\times 10^{-5}$ & $9.89 \times 10^{-5}$ \\
 $a_{1}$  &   $3.68\times 10^{-6}$ &$ -4.96\times 10^{-6}$ & $ -1.46\times 10^{-5}$ \\
 $b_{1}$  &   $ 1.95\times 10^{-6}$ & $4.55\times 10^{-6}$  & $ -2.34\times 10^{-5} $ \\
 $a_{2}$  &  $  3.70\times 10^{-6}$ &  ---  &  ---  \\
 $b_{2}$  &  $8.68\times 10^{-7}$ &   ---  &  ---  \\
 $\omega$   &   $  0.0250  $   &$ 0.0250 $ &$ 0.0575 $ \\
\hline
RMS Error   &  $9.77\times 10^{-7} $ & $1.84\times 10^{-6}$ &  $7.29\times 10^{-6}$  \\   
\hline \\
\end{tabular}
\end{table}

Once trained, the parameter functions that are found can be tested, by using the Hamiltonian so defined to calculate the QNN indicator for  other initial states. Testing was therefore done on a large number of states not represented in the training set, including fully entangled states, partially entangled states, product (unentangled) states, and even \underline{mixed} states. (Note that only \underline{pure} states were present in the training sets.) The interested reader is referred to our previous work for the (extensive) testing results \cite{behrmanqic, behrmanmulti, behrmanieee}. Note also that the testing was done using the fitted functions only.

\section{ Learning with noise} 

Physical systems contain noise, meaning that there is some uncertainty in each of the elements of the density matrix (though, of course, it must remain hermitian and positive semidefinite, with unit trace to conserve probability.) What if the system on which we train is somewhat noisy? 

Let us first define our terms. Because we are working with simulations, we can isolate the different effects of ``noise'' and ``decoherence'': here, we will use ``noise'' to refer to random (uncorrelated) magnitudes, of a given size, added to the density matrix elements; and ``decoherence'' to refer to random phases so added. In general, of course, both effects will be present; we will call that ``complex'' noise. Recall that our entanglement indicator is a mapping to a time correlation function, after evolution in time according to the Hamiltonian of Equation \ref{ham}; that is, the entanglement of the initial state is approximated by a measurement performed at the final time. To simulate white noise (zero mean and specified amplitude), random numbers were added at each timestep $\Delta t=0.8$ ns of that time evolution. These numbers have zero time correlation themselves. The level of noise we report is the amplitude, that is, the root-mean-square-average size of these random numbers, that are imposed at each time step. All the simulations were done for the same total time of 317 timesteps, or about 253 ns. Because the system evolves in time, the noise itself propagates, so that by the time the correlation function is measured, at the final time, numbers that seem quite small can build up to destroy a significant amount of entanglement. For comparison with our testing results, we will always therefore include the entanglement of formation for the noisy density matrix, calculated using the Bennett-Wootters formula\cite{bennett2,wootters} and marked ``BW'' on our testing graphs (Figures 12-17, 25-30, and 38-43) below.

%\begin{figure}
%\includegraphics[height=2in]{errorfornonzeron.eps} 
%\fcaption{Error for the training set of Table~\ref{enttraining}, as a function of noise level. \label{errnon0n}}
%\end{figure}

\begin{figure}
\includegraphics[height=2in]{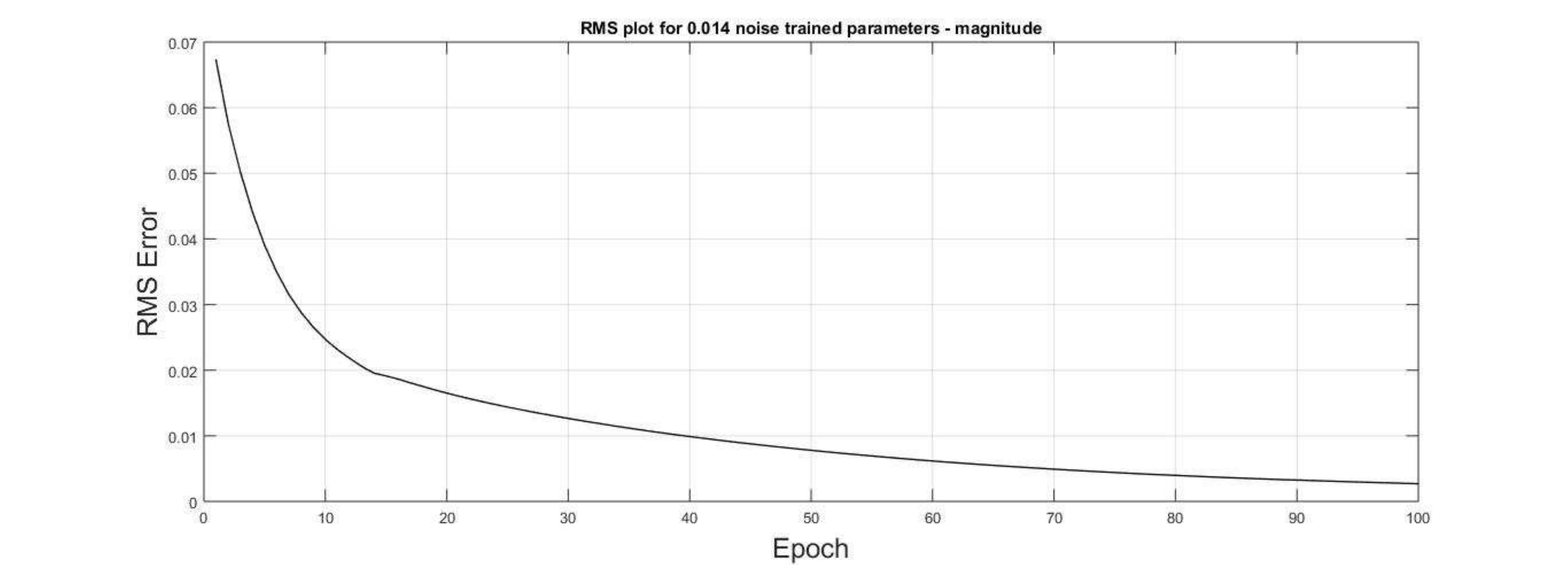} 
\caption{Total root mean squared error for the training set as a function of epoch (pass through the training set), for the 2-qubit system, with a (magnitude) noise level of 0.014 at each (of 317 total) timestep. Asymptotic error is $3.1 \times 10^{-3}$, about double what it was with no noise.}
\label{magtrn}
\end{figure}

\begin{figure}
\includegraphics[height=2in]{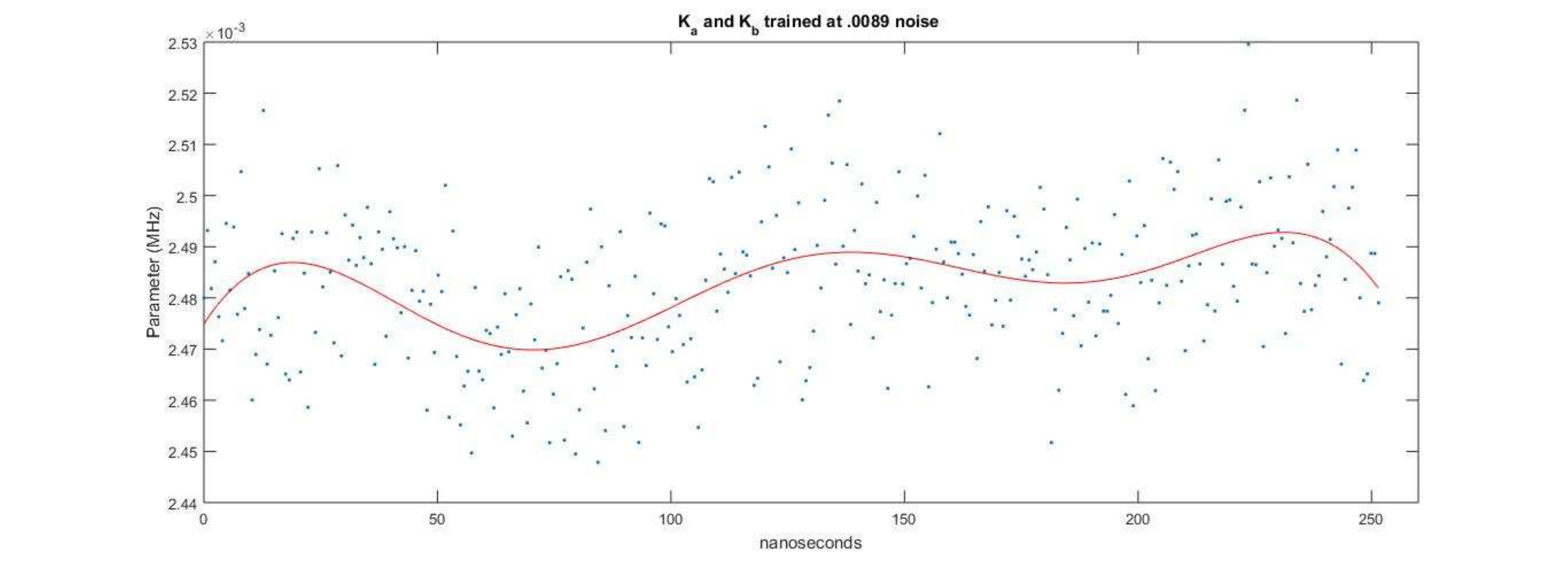} 
\caption{Parameter function $K_{A}=K_{B}$ as a function of time, as trained at 0.0089 amplitude noise at each of the 317 timesteps, for the entanglement indicator (data points), and plotted with the Fourier fit (solid line). Note the change in scale from Figure~\ref{K0}, because of the (much larger) spread of the noisy data: the Fourier fit is actually almost the same on this graph.}
\label{kmag}
\end{figure}

\begin{figure}
\includegraphics[height=2in]{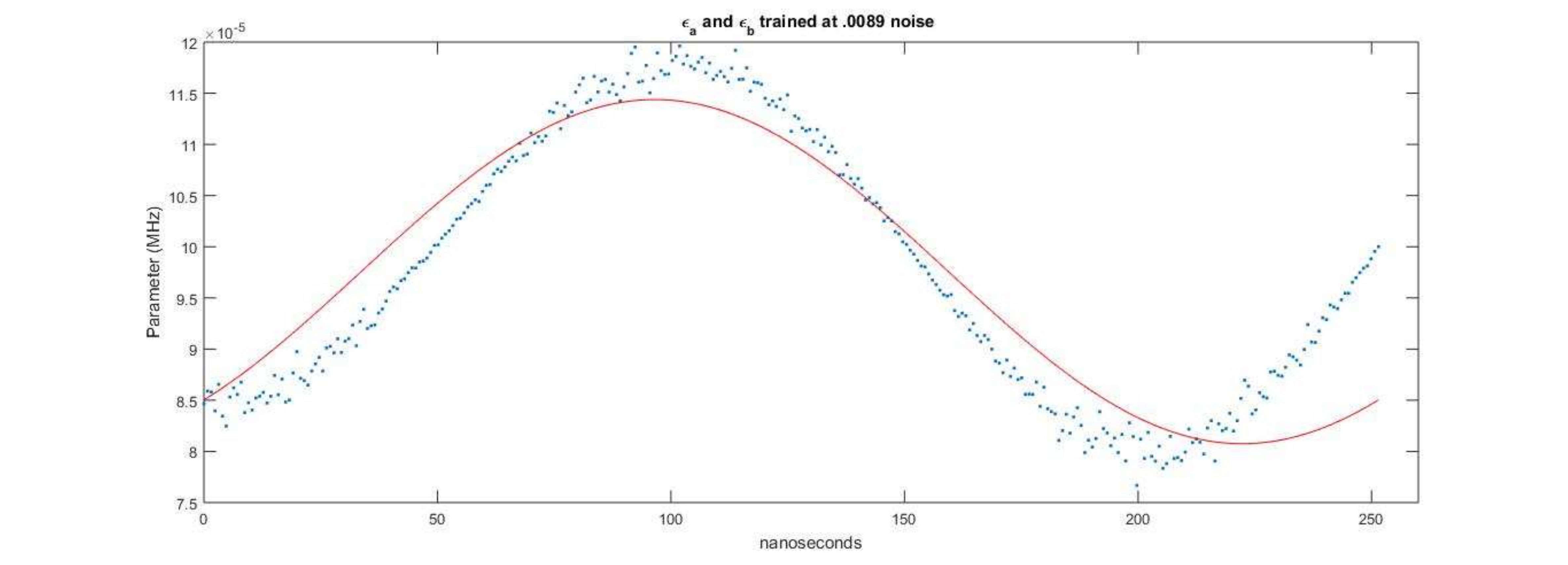} 
\caption{Parameter function $\epsilon_{A}=\epsilon_{B}$ as a function of time, as trained at 0.0089 amplitude noise at each of the 317 timesteps,  for the entanglement indicator (data points), and plotted with the Fourier fit (solid line).}
\label{epmag}
\end{figure}

\begin{figure}
\includegraphics[height=2in]{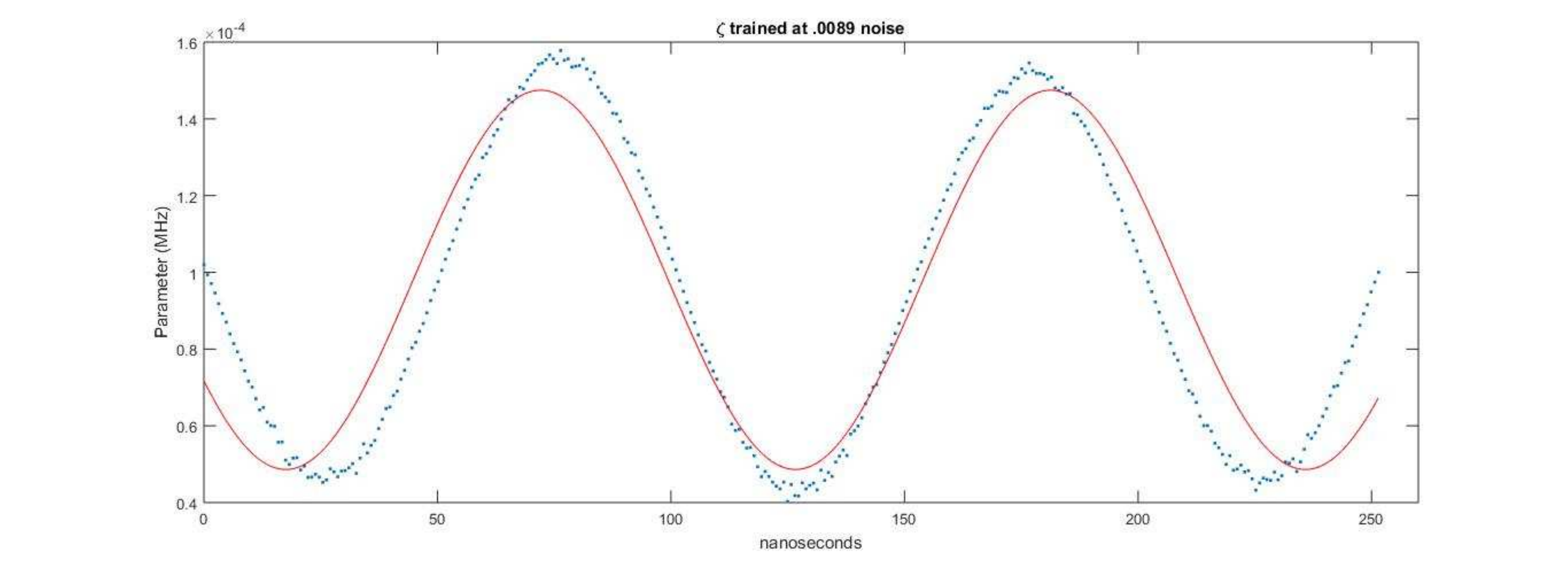} 
\caption{Parameter function $\zeta$ as a function of time, as trained at 0.0089  noise at each of the 317 timesteps,  for the entanglement indicator (data points), and plotted with the Fourier fit (solid line).}
\label{zmag}
\end{figure}

We consider first the case of  magnitude noise only. Figure~\ref{magtrn} shows a typical rms error training curve for a fairly large level of noise. As expected, asymptotic error for the training set does increase with increasing noise. The parameter functions also become ``noisy'': See Figures~\ref{kmag},\ref{epmag},\ref{zmag}. 


Much of this variation is not meaningful, though. In fact, the numbers in Table~\ref{entparamfit} are slightly different from the ones we found in \cite{behrmanieee}. Is that difference significant? To investigate this, we tried testing with all but a selected one of the parameter functions' Fourier coefficients set to the trained values, but assigning random numbers (of the right order of magnitude) to the Fourier coefficients of the one chosen. The system was remarkably insensitive to this procedure when the random function was the tunneling function $K$: as long as $K$ was the right order of magnitude, the indicator still tested extremely well - in fact for most of the testing the indicator results were identical. This was not true if $\epsilon$ or $\zeta$ were randomized, however: errors were substantial, particularly when the Fourier frequency $\omega$ is randomized. Still, exact agreement with the trained functions is not, apparently, necessary. 

\begin{figure}
\includegraphics[height=2in]{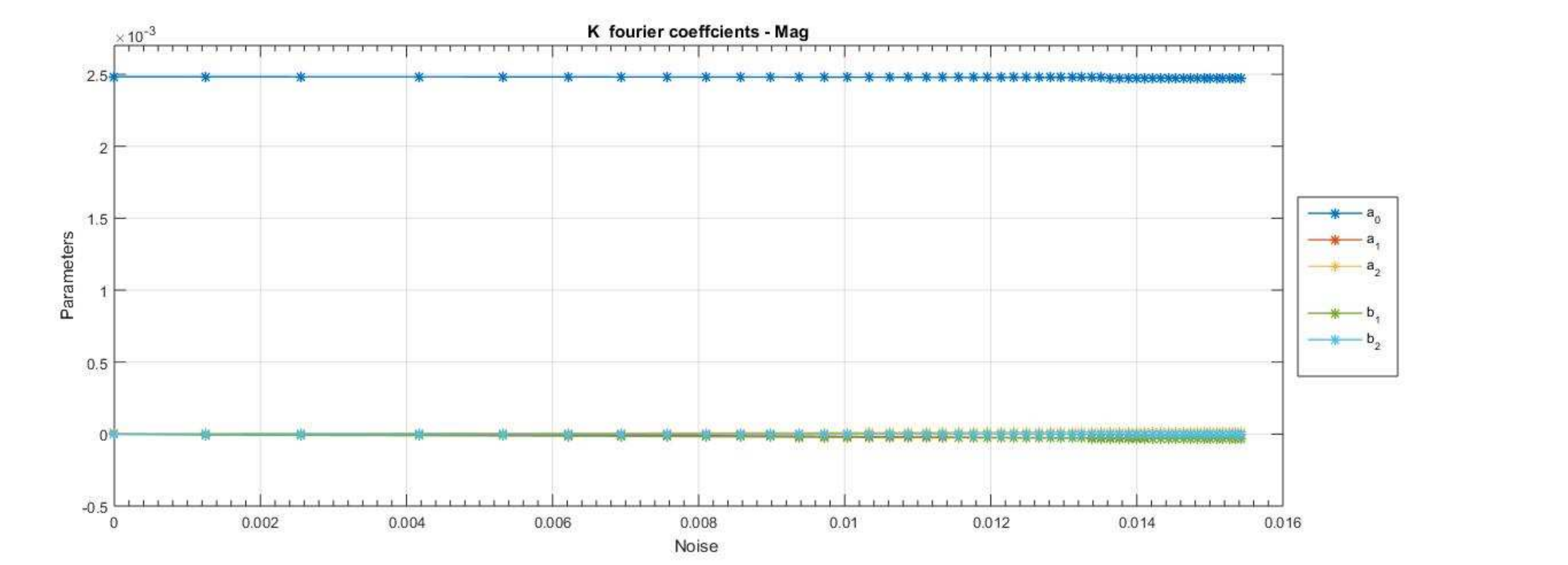} 
\caption{Fourier coefficients for the tunneling parameter functions $K$, as functions of noise level.}
\label{FKmag}
\end{figure}

\begin{figure}
\includegraphics[height=2in]{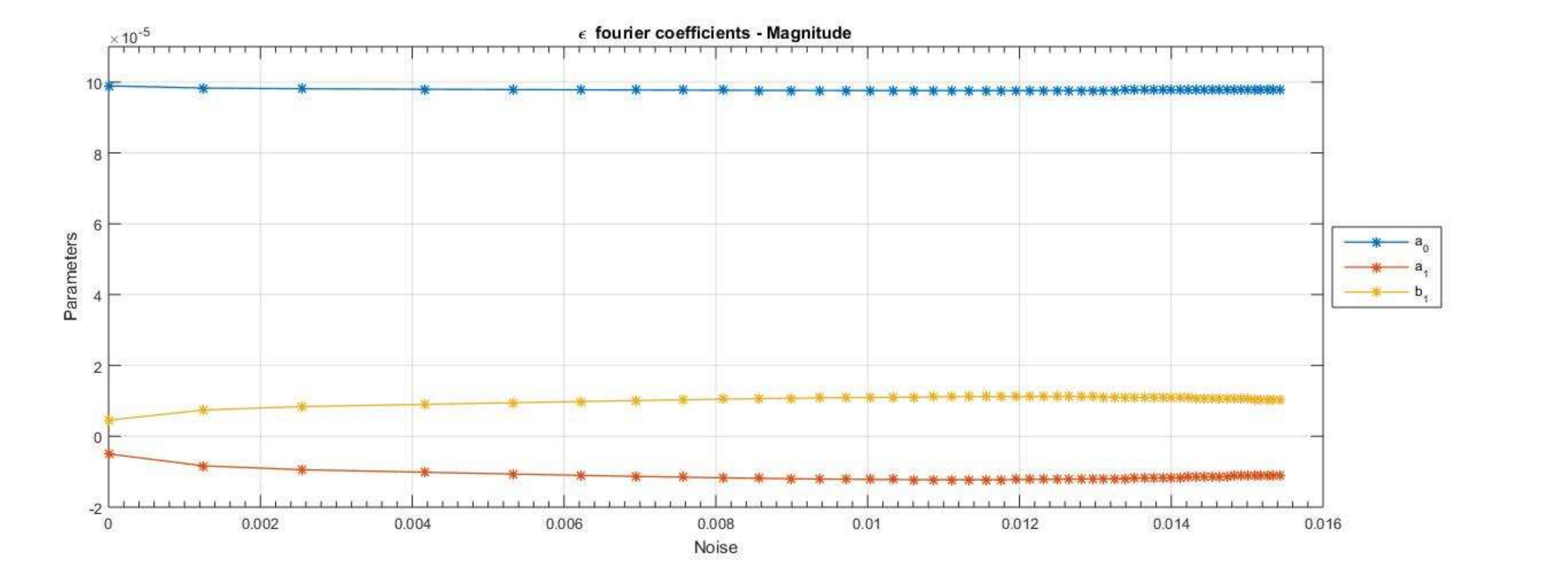} 
\caption{Fourier coefficients for the bias parameter functions $\epsilon$, as functions of noise level. }
\label{Fepmag}
\end{figure}

\begin{figure}
\includegraphics[height=2in]{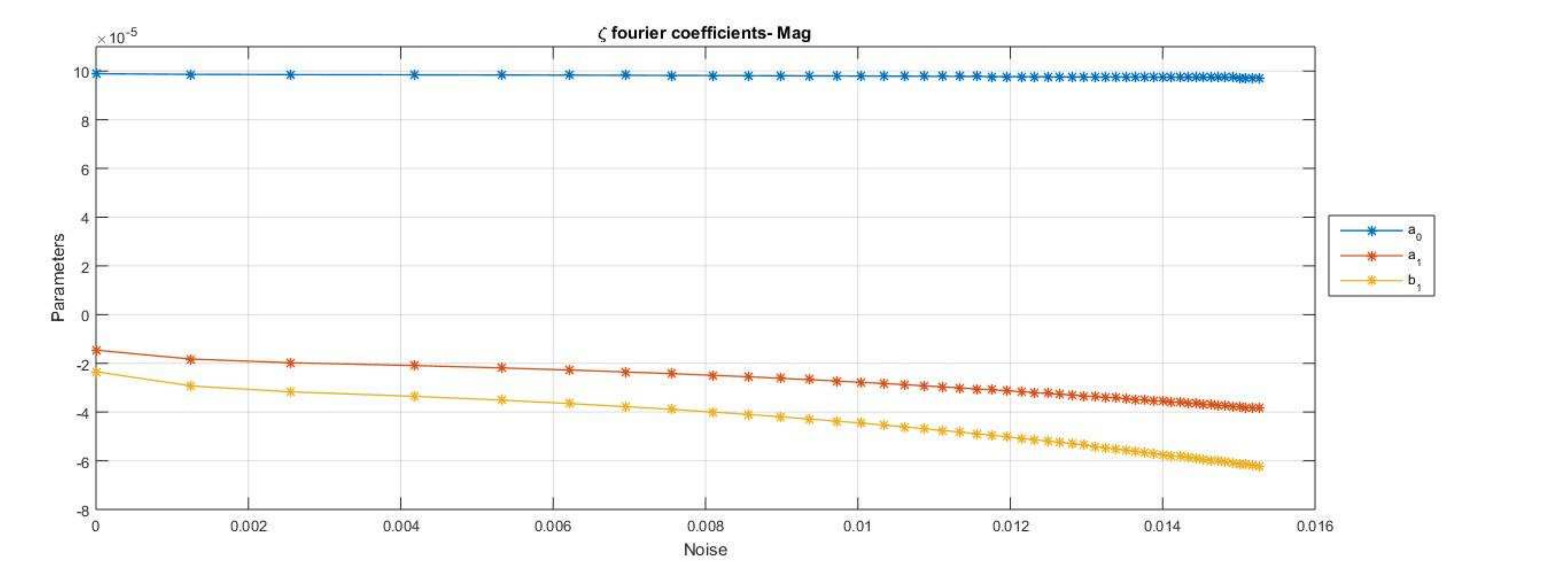} 
\caption{Fourier coefficients for the coupling parameter function $\zeta$, as functions of noise level.}
\label{Fzmag}
\end{figure}

We can, as before, fit each function to a Fourier series, and test using the fitted functions. Figures~\ref{FKmag},\ref{Fepmag},\ref{Fzmag} show the Fourier coefficients for $K_{A}=K_{B}$, $\epsilon_{A}=\epsilon_{B}$ , and $\zeta$, respectively, as functions of increasing noise level. The Fourier components of the tunneling parameter function $K$ are clearly the least sensitive to noise level, while for $\epsilon$ they are a bit more so and for $\zeta$ the most. This is in accordance with the observed insensitivity of the entanglement indicator to $K$: that is, it is true both that the indicator is relatively insensitive to the $K$ function, and that the $K$ function's Fourier fit is insensitive to environmental noise.


All of the parameters look fairly stable at these levels of noise, from this point of view. But a more important question is: how much does noise interfere with the net's ability to detect entanglement? 

\begin{figure}
\includegraphics[height=2in]{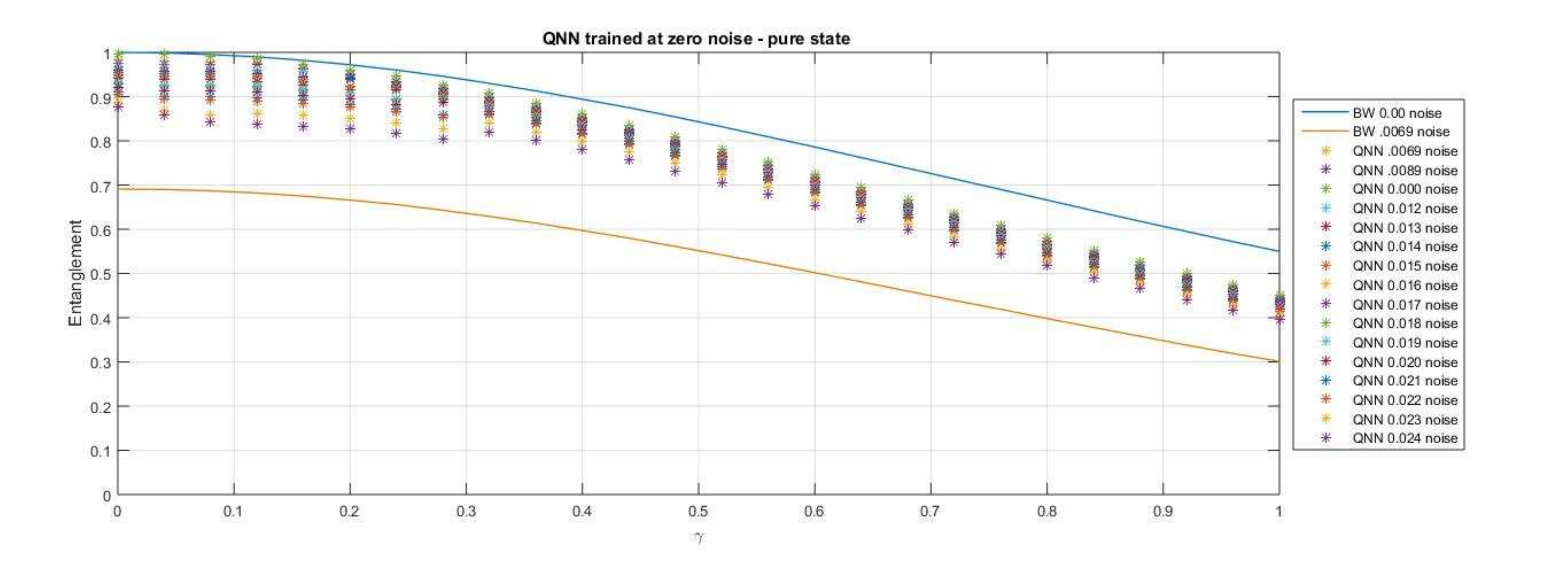} 
\caption{Entanglement of the state $P$ as a function of $\gamma$, as calculated by the QNN, and compared with the entanglement of formation (marked ``BW'') at zero noise (blue) and at 0.0069 noise (orange). In each case the QNN was trained at zero noise, but tested at the given level.}
\label{P0mag}
\end{figure}

\begin{figure}
\includegraphics[height=2in]{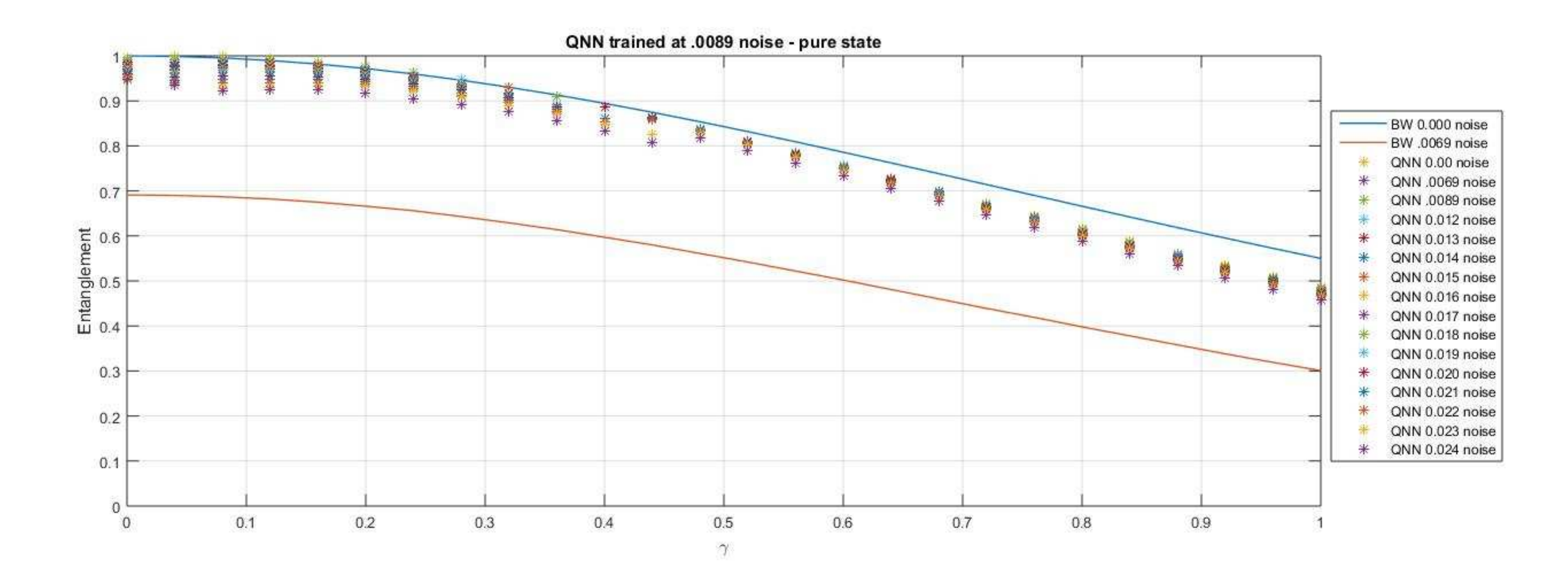} 
\caption{Entanglement of the state $P$ as a function of $\gamma$, as calculated by the QNN, and compared with the entanglement of formation (marked ``BW'') at zero noise (blue) and at 0.0069 noise (orange). In each case the QNN was trained at a noise level of 0.0089, and then tested at the given level.}
\label{P89mag}
\end{figure}

\begin{figure}
\includegraphics[height=2in]{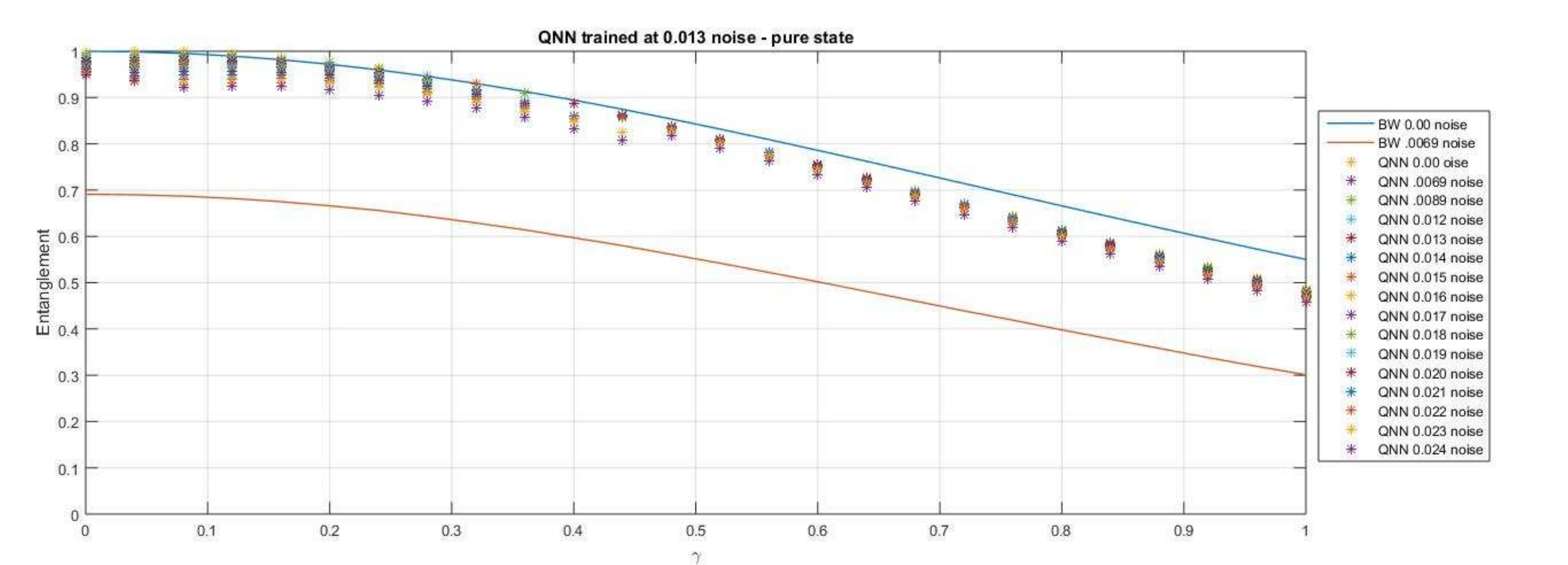} 
\caption{Entanglement of the state $P$ as a function of $\gamma$, as calculated by the QNN, and compared with the entanglement of formation (marked ``BW'') at zero noise (blue) and at 0.0069 noise (orange). In each case the QNN was trained at 0.013 noise, and then tested at the given level.}
\label{P13mag}
\end{figure}

Consider a pure state, specifically, the state $P(\gamma) = \frac{|00 \rangle + |11 \rangle + \gamma |01 \rangle}{\sqrt{2+|\gamma|^{2}}}$. For $\gamma=0$ this is of course the (fully entangled) Bell state; as $\gamma$ increases the entanglement decreases. $P(\gamma)$ is one of the states we tested on in the 2008 paper\cite{behrmanqic}, and, as we found then, the entanglement as computed by the quantum neural network, without noise, tracks the entanglement of formation very well.  How much noise or uncertainty can the network tolerate? We answer this question in two ways. First, we suppose the QNN was trained on the perfect (zero noise) system of the four-pair training set, and we add increasing amounts of noise to test the system for various nonzero values of $\gamma$, and compare the results to the entanglement of formation both at zero noise and at 0.0069 noise. See Figure~\ref{P0mag}. Then, we suppose the QNN was trained on a noisy system, and, again, test with increasing levels of noise. Figures~\ref{P89mag} and \ref{P13mag} show, respectively, an intermediate level and a high level of noise present during training. Recall that the noise should be understood as occurring over the total time evolution interval: that is, an independent (uncorrelated) noise at the given rms level was added at \underline{each} timestep. The two curves showing the entanglement of formation can therefore be thought of as a kind of ``error bars'': the correct entanglement of the system should lie somewhere between the zero noise result and the result at maximum noise, insofar as the QNN tracks well with the entanglement of formation. Presumably the presence of noise does destroy entanglement, but, since the measurement itself is noisy, it is not certain how much is destroyed and how much is simply a bad measurement. Still, it is obvious from the results that, indeed, the QNN technique does an excellent job of remaining robust to pure noise.


\begin{figure}
\includegraphics[height=2in]{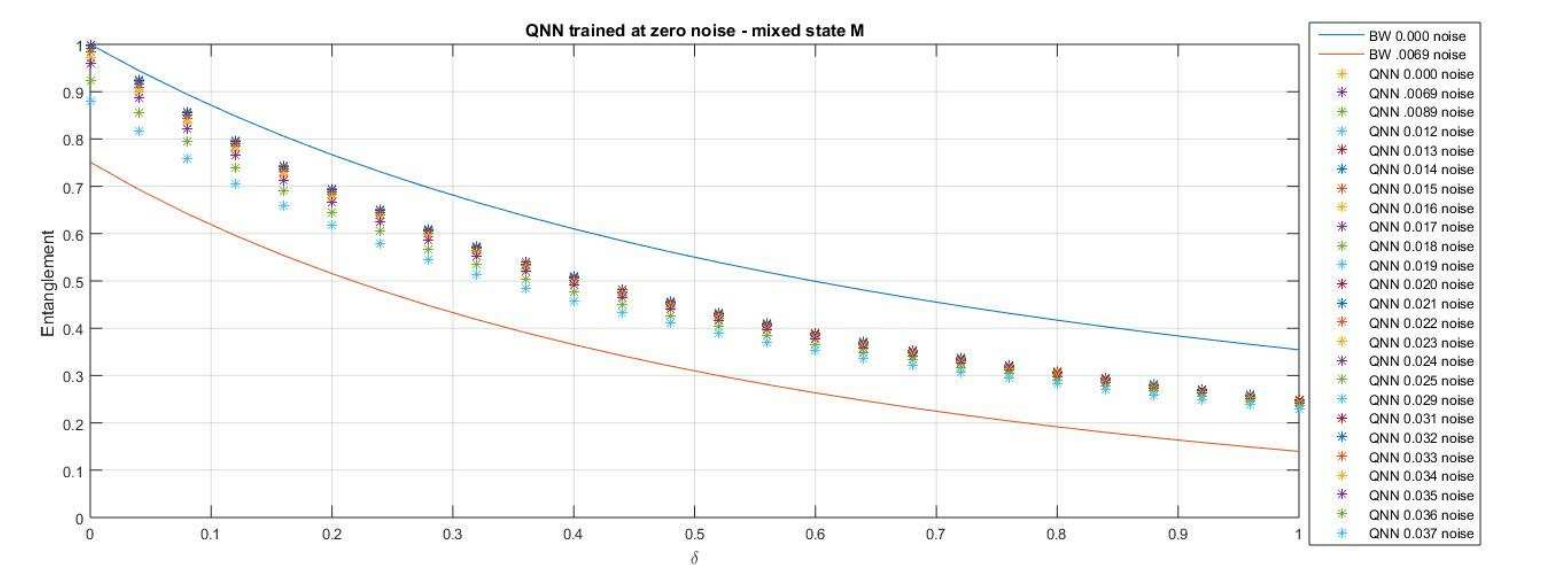} 
\caption{Entanglement of the state $M$ as a function of $\delta$, as calculated by the QNN, and compared with the entanglement of formation (marked ``BW'') at zero noise (blue) and at 0.0069 noise (orange). In each case the QNN was trained at zero noise, but tested at the given level.}
\label{M0mag}
\end{figure}

\begin{figure}
\includegraphics[height=2in]{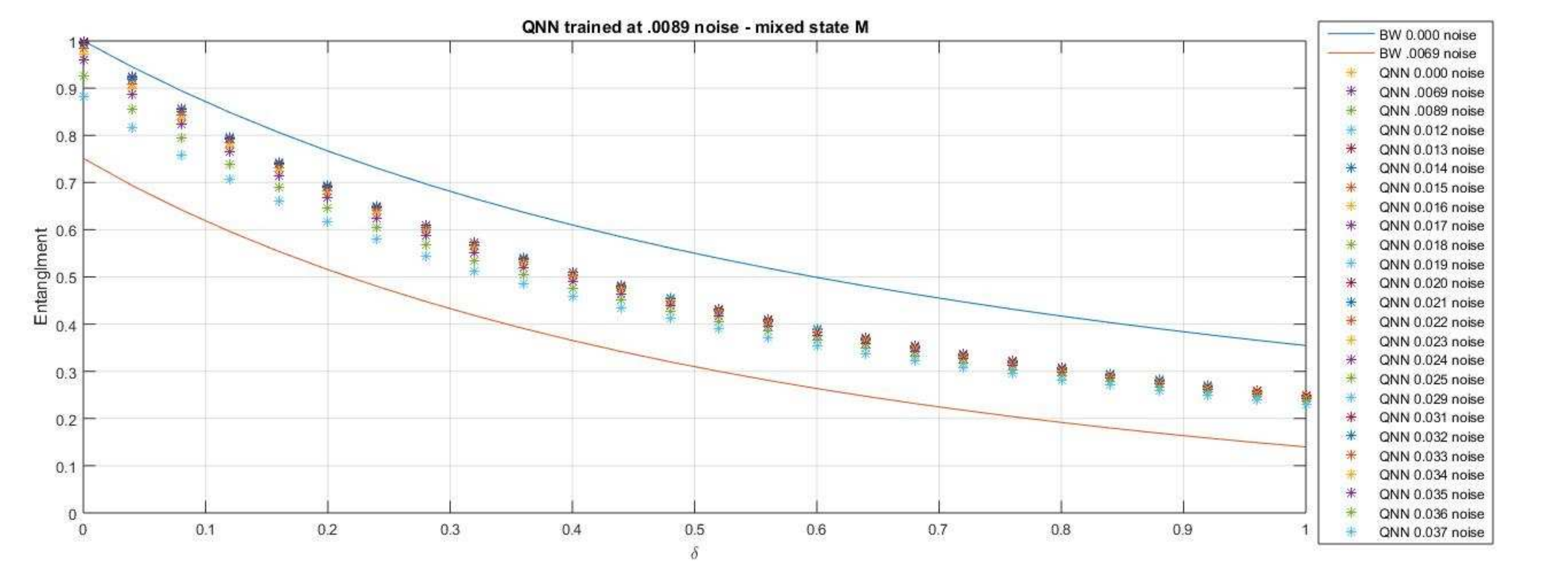} 
\caption{Entanglement of the state $M$ as a function of $\delta$, as calculated by the QNN, and compared with the entanglement of formation (marked ``BW'') at zero noise (blue) and at 0.0069 noise (orange). In each case the QNN was trained at a noise level of 0.0089, and then tested at the given level.}
\label{M89mag}
\end{figure}

\begin{figure}
\includegraphics[height=2in]{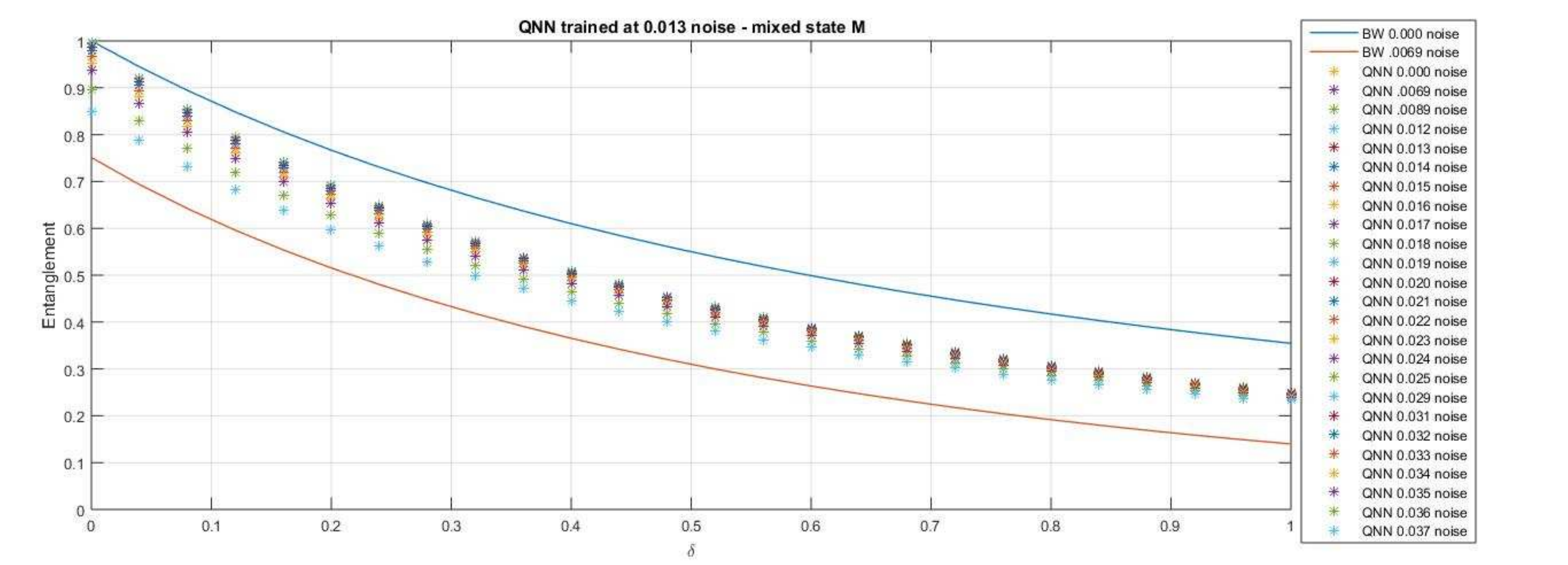} 
\caption{Entanglement of the state $M$ as a function of $\delta$, as calculated by the QNN, and compared with the entanglement of formation (marked ``BW'') at zero noise (blue) and at 0.0069 noise (orange). In each case the QNN was trained at 0.013 noise, and then tested at the given level.}
\label{M13mag}
\end{figure}

Second, we consider testing on mixed states. We might expect the QNN to perform significantly less well with these kinds of states, because the training set (Table \ref{enttraining}) contained no mixed states. In fact with zero noise the QNN tested well on several classes of mixed states \cite{behrmanqic}; we need to see if that success is maintained with noisy conditions. Figures~\ref{M0mag},\ref{M89mag},\ref{M13mag} show results for the mixed states $M(\delta) = \frac{\delta|11 \rangle \langle 11| + |\Phi^{+} \rangle \langle \Phi^{+} |}{\delta + 1}$, where $|\Phi^{+} \rangle $ is the Bell state, given in Table \ref{enttraining}, as functions of $\delta$, for QNN trained at zero, 0.0089, and 0.013 noise amplitude, respectively.  Again, the entanglement of formation results can serve as an approximate error bound, and we see that the QNN's entanglement indicator is robust to noise. Indeed, comparison of these figures with the ones for pure states shows that performance on mixed states is even better. Because of this we might expect the QNN indicator to show greater robustness to decoherence than to magnitude noise; we will see in the next section that this is, indeed, the case.


\section{ Decoherence} 

\begin{figure}
\includegraphics[height=2in]{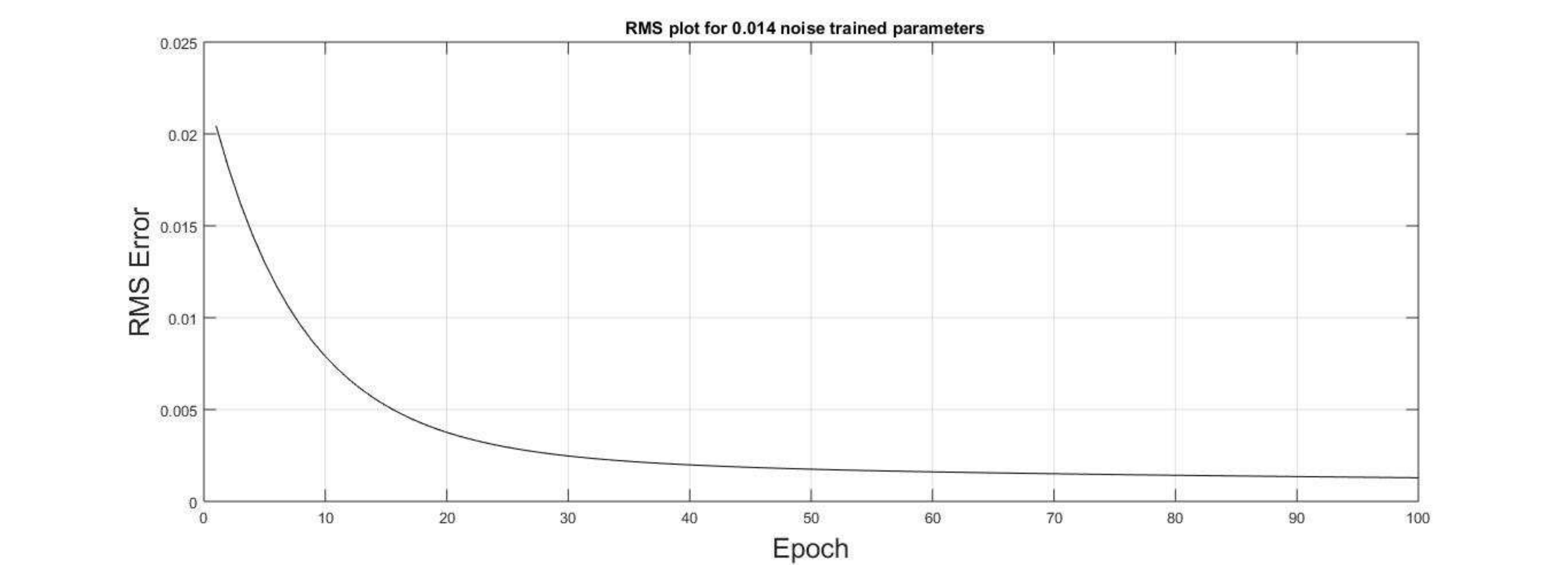} 
\caption{Total root mean squared error for the training set as a function of epoch (pass through the training set), for the 2-qubit system, with a phase noise level of 0.014 at each (of 317 total) timestep. Asymptotic error is $1.3 \times 10^{-3}$, approximately the same as with no noise.}
\label{phtrn}
\end{figure}

\begin{figure}
\includegraphics[height=2in]{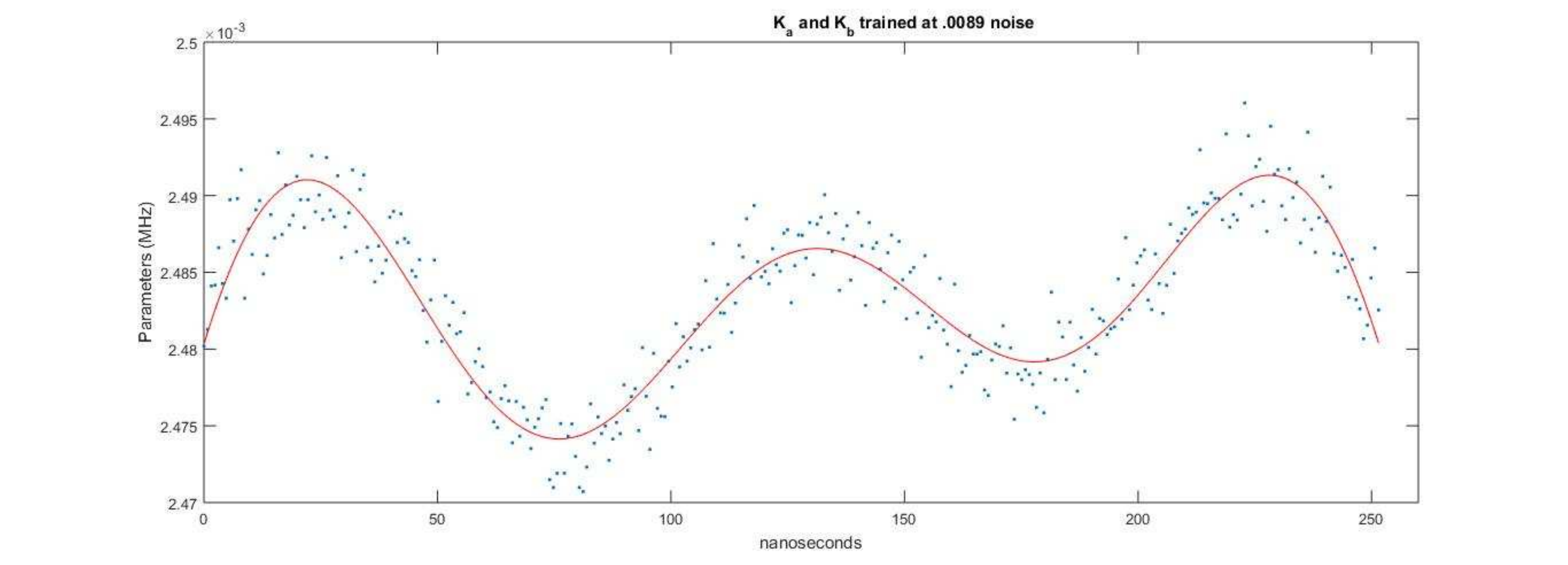} 
\caption{Parameter function $K_{A}=K_{B}$ as a function of time, as trained at 0.0089 phase noise at each of the 317 timesteps, for the entanglement indicator (data points), and plotted with the Fourier fit (solid line)}
\label{Kph}
\end{figure}

\begin{figure}
\includegraphics[height=2in]{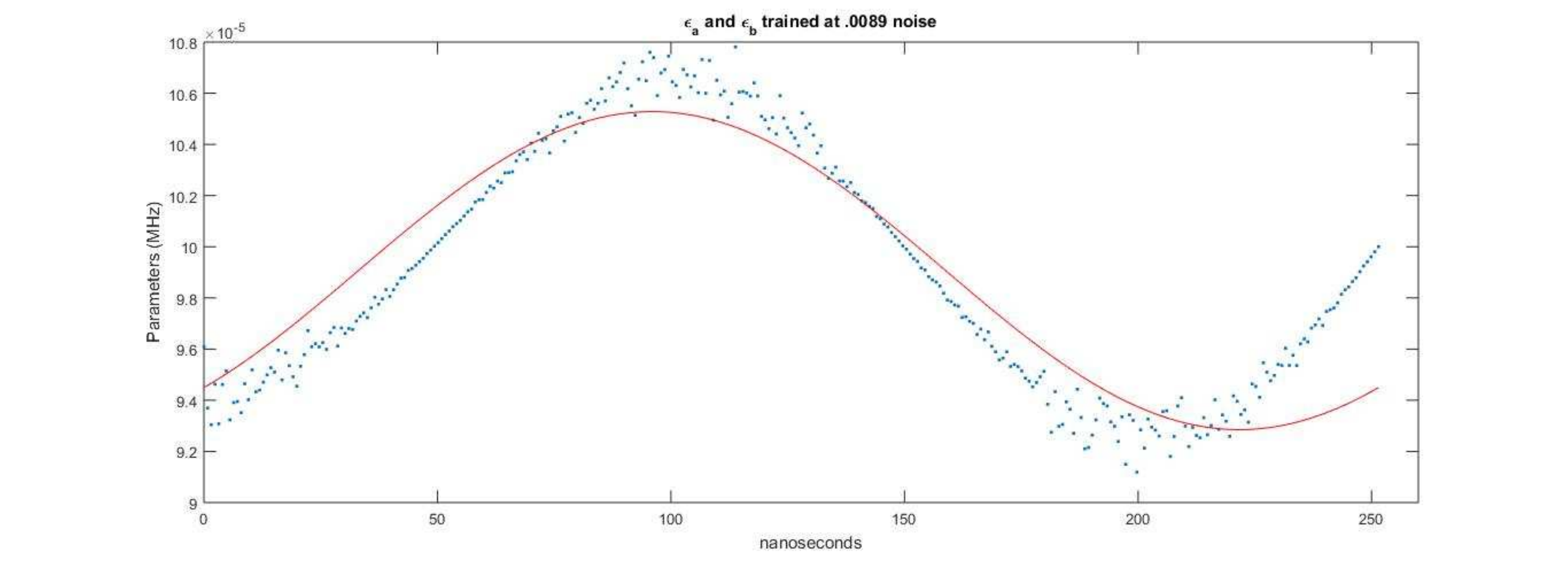} 
\caption{Parameter function $\epsilon_{A}=\epsilon_{B}$ as a function of time, as trained at 0.0089 phase noise at each of the 317 timesteps,  for the entanglement indicator (data points), and plotted with the Fourier fit (solid line).}
\label{epph}
\end{figure}

\begin{figure}
\includegraphics[height=2in]{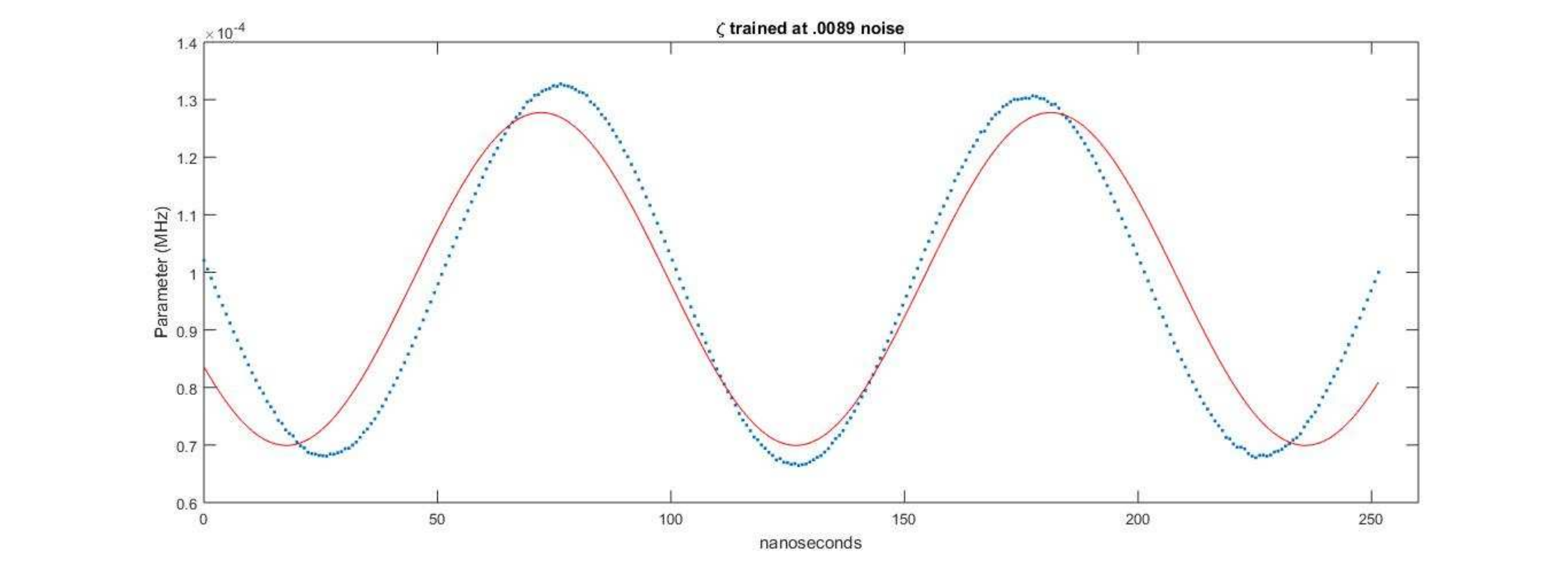} 
\caption{Parameter function $\zeta$ as a function of time, as trained at 0.0089 phase noise at each of the 317 timesteps,  for the entanglement indicator (data points), and plotted with the Fourier fit (solid line).}
\label{zph}
\end{figure}

We now turn to the case of ``pure'' decoherence, that is, we introduce random phases to the elements of the density matrix, without changing their magnitudes. Figure~\ref{phtrn} shows a typical rms error training curve for a fairly large level of phase noise. Again, asymptotic error for the training set does increase with increasing phase noise (decoherence). The parameter functions also become ``noisy'': See Figures~\ref{Kph},\ref{epph},\ref{zph}.


\begin{figure}
\includegraphics[height=2in]{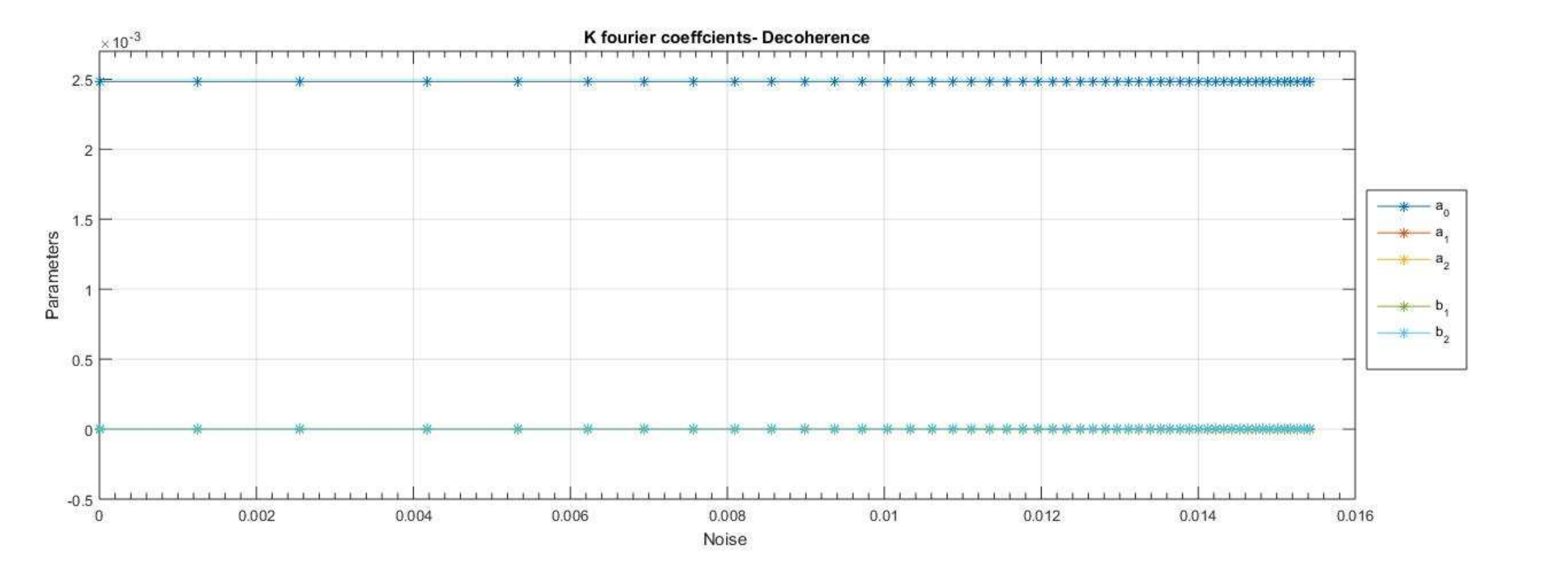} 
\caption{Fourier coefficients for the tunneling parameter functions $K$, as functions of decoherence level.}
\label{FKph}
\end{figure}

\begin{figure}
\includegraphics[height=2in]{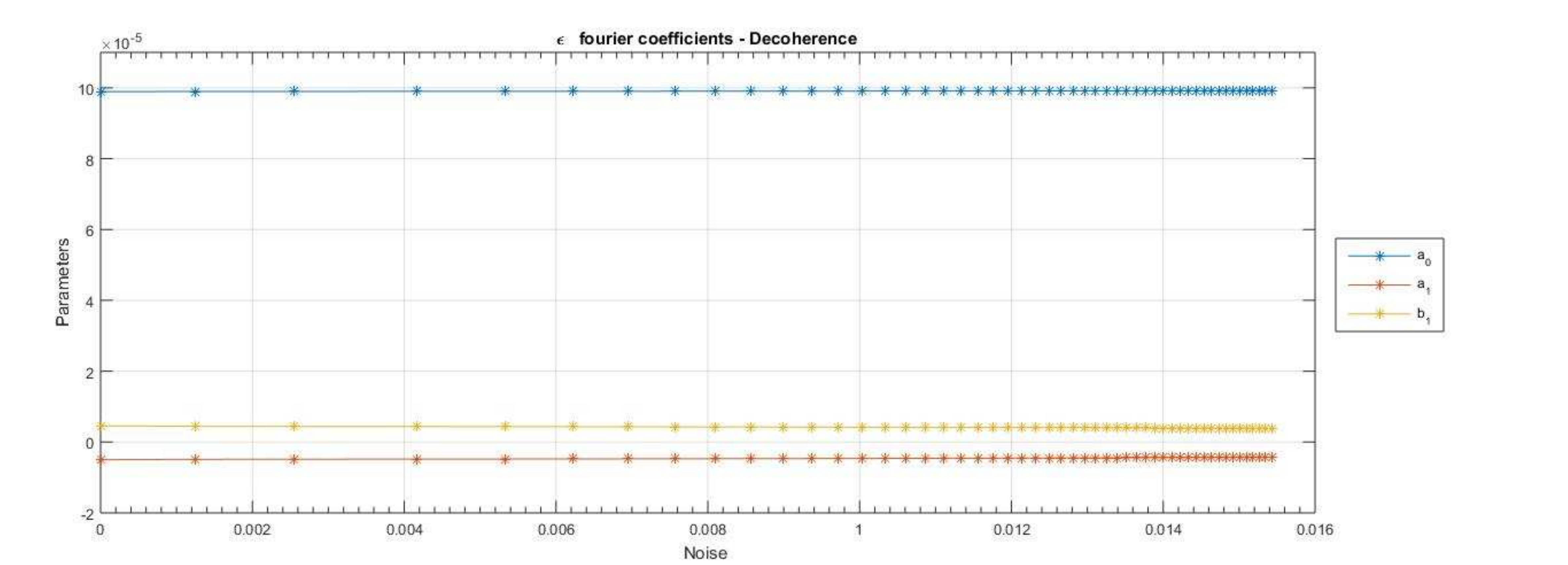} 
\caption{Fourier coefficients for the bias parameter functions $\epsilon$, as functions of decoherence level.}
\label{Fepph}
\end{figure}

\begin{figure}
\includegraphics[height=2in]{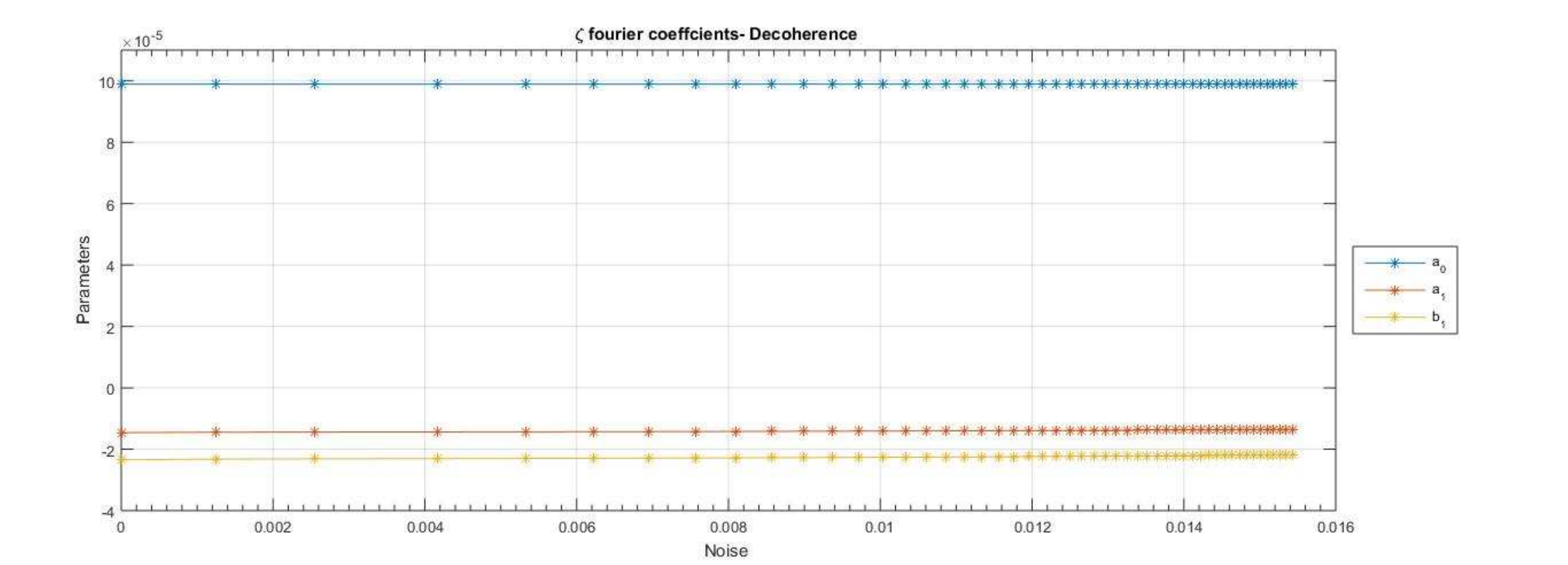} 
\caption{Fourier coefficients for the coupling parameter function $\zeta$, as functions of decoherence level.}
\label{Fzph}
\end{figure}

Figures~\ref{FKph},\ref{Fepph},\ref{Fzph} show the Fourier coefficients for $K_{A}=K_{B}$, $\epsilon_{A}=\epsilon_{B}$ , and $\zeta$, respectively, as functions of increasing decoherence level. It is clear that in the case of decoherence the parameter functions are even more stable than in the previous case of noise.


\begin{figure}
\includegraphics[height=2in]{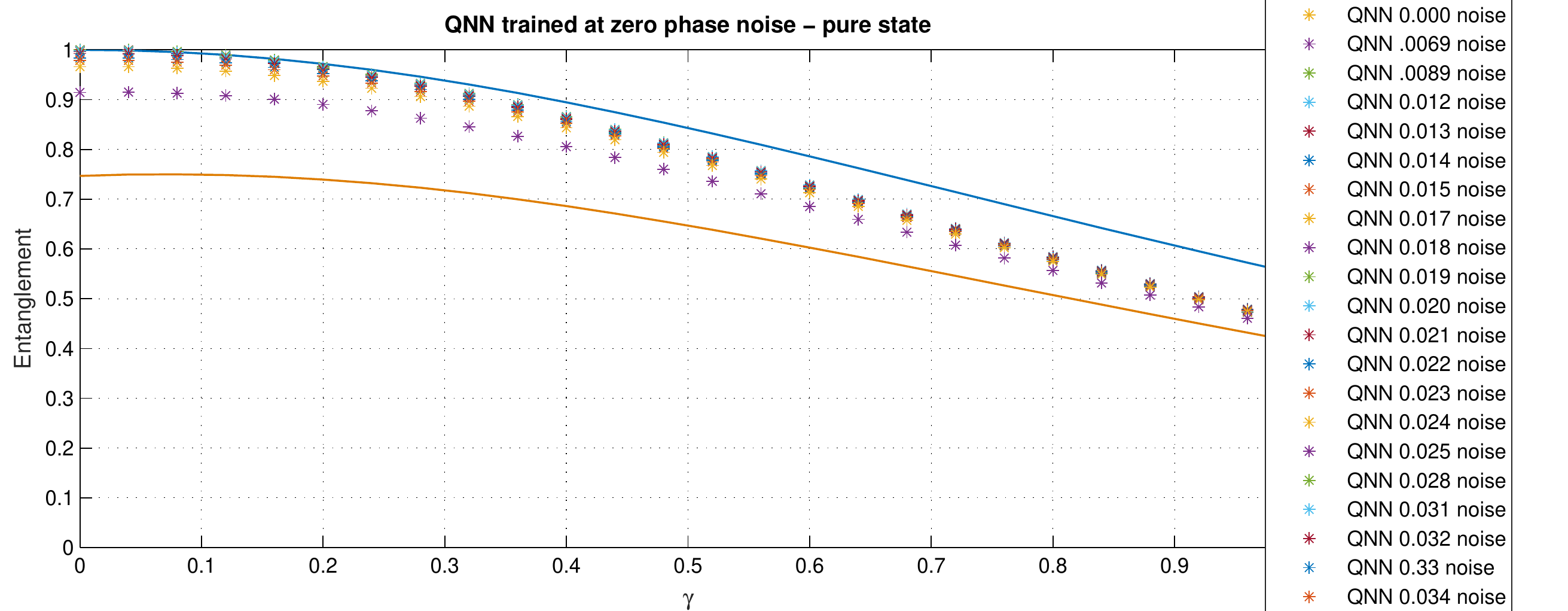} 
\caption{Entanglement of the state $P$ as a function of $\gamma$, as calculated by the QNN, and compared with the entanglement of formation (marked ``BW'') at zero phase noise (blue) and at 0.0069 phase noise (orange). In each case the QNN was trained at zero phase noise, but tested at the given level.}
\label{P0ph}
\end{figure}

\begin{figure}
\includegraphics[height=2in]{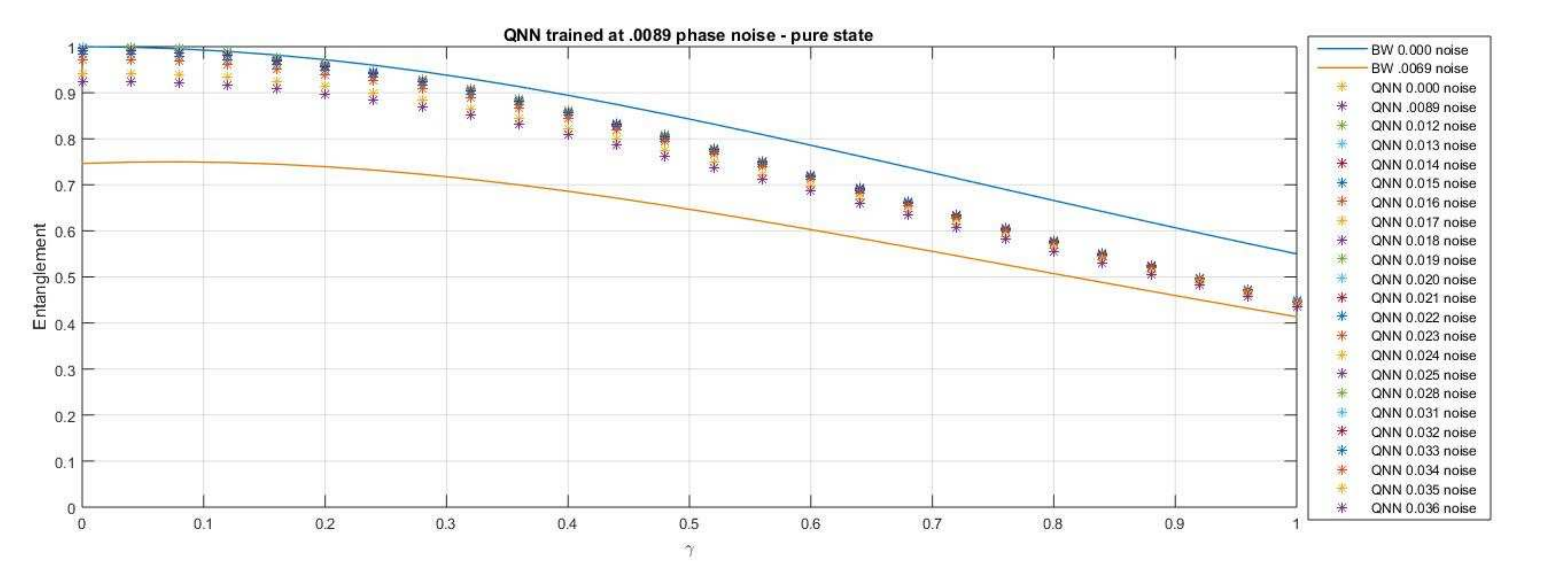} 
\caption{Entanglement of the state $P$ as a function of $\gamma$, as calculated by the QNN, and compared with the entanglement of formation (marked ``BW'') at zero phase noise (blue) and at 0.0069 phase noise (orange). In each case the QNN was trained at a phase noise level of 0.0089, and then tested at the given level.}
\label{P89ph}
\end{figure}

\begin{figure}
\includegraphics[height=2in]{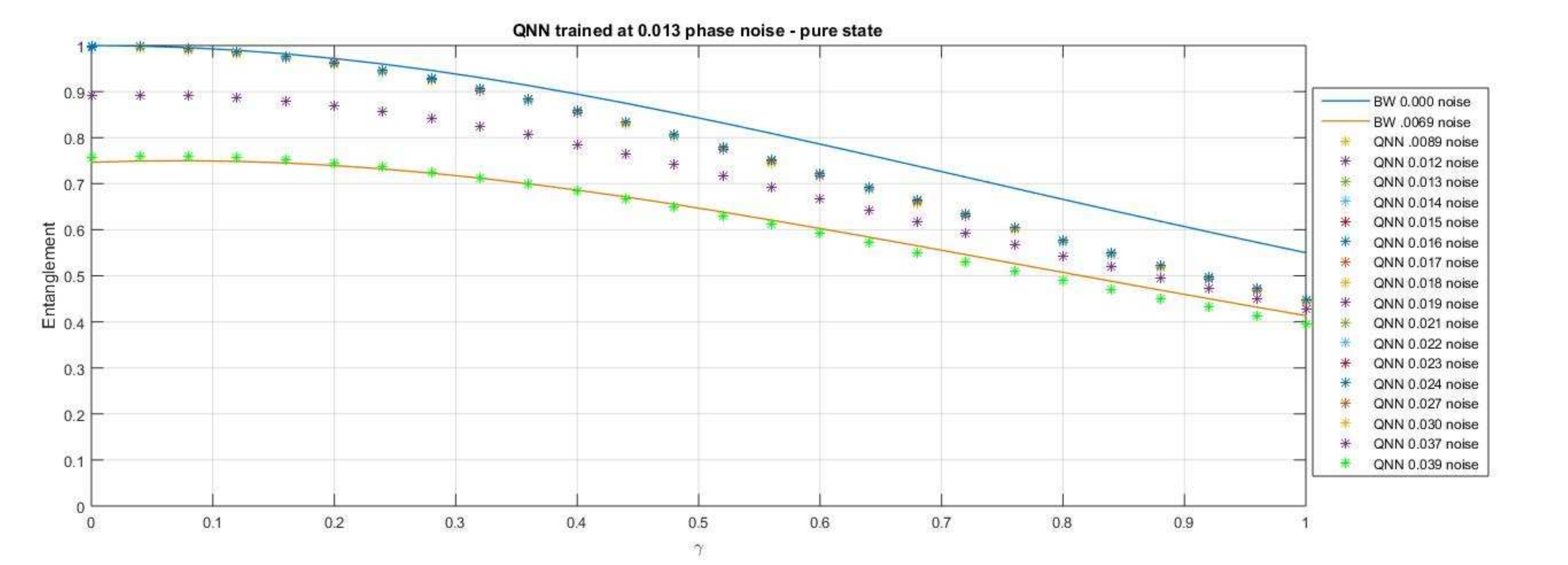} 
\caption{Entanglement of the state $P$ as a function of $\gamma$, as calculated by the QNN, and compared with the entanglement of formation (marked ``BW'') at zero phase noise (blue) and at 0.0069 phase noise (orange). In each case the QNN was trained at 0.013 phase noise, and then tested at the given level.}
\label{P13ph}
\end{figure}

Testing results for the pure state $P$ subject to decoherence are shown in Figures~\ref{P0ph},\ref{P89ph},\ref{P13ph}, as trained, respectively, at zero, 0.0089, and 0.013 decoherence (phase noise), and tested with various levels of phase noise, and compared with the entanglement of formation at zero and at 0.0069 phase noise. The results are extremely good, better even than the ones in Figures~\ref{P0mag},\ref{P89mag},\ref{P13mag}; clearly, the QNN is even better at dealing with decoherence than with ``pure'' noise. Possibly the various (random) phases tend to cancel each other; but since for \underline{definite} phase shifts the QNN underestimates the entanglement, sometimes drastically \cite{behrmanqic,behrmanieee}, this was an unexpectedly good result.


\begin{figure}
\includegraphics[height=2in]{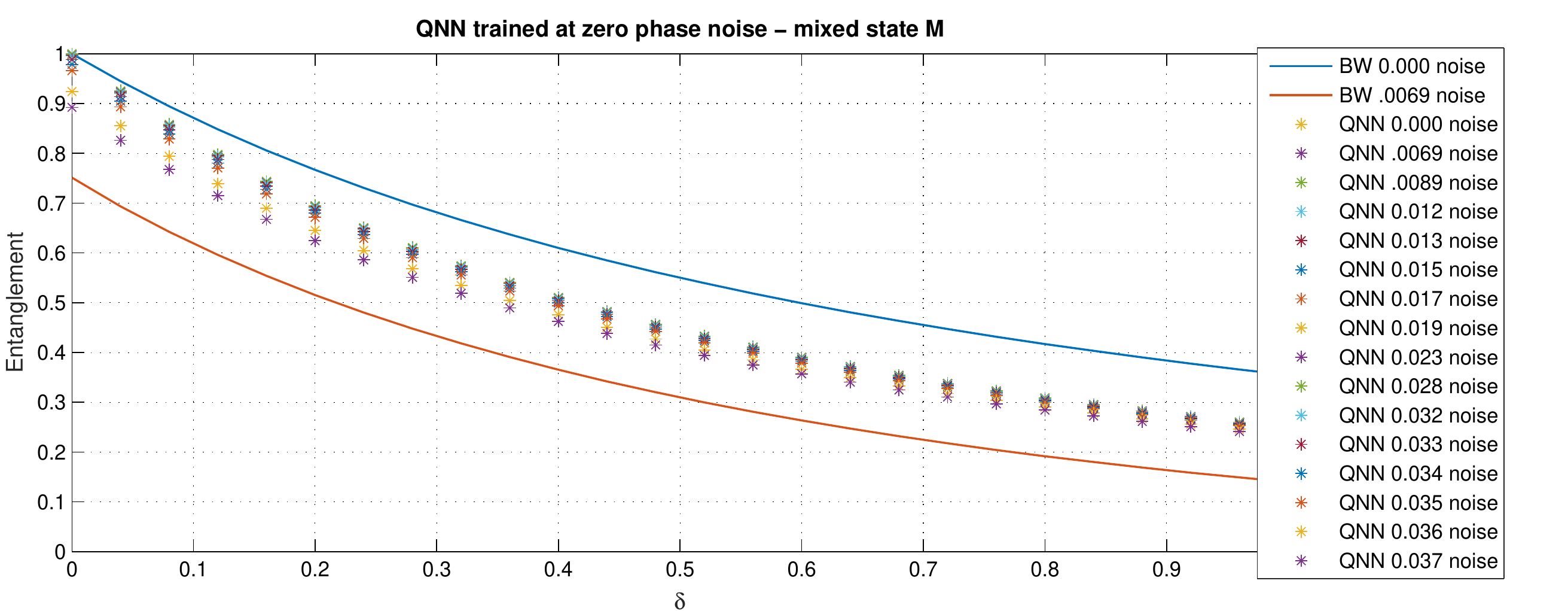} 
\caption{Entanglement of the state $M$ as a function of $\delta$, as calculated by the QNN, and compared with the entanglement of formation (marked ``BW'') at zero phase noise (blue) and at 0.0069 phase noise (orange). In each case the QNN was trained at zero noise, but tested at the given level.}
\label{M0ph}
\end{figure}

\begin{figure}
\includegraphics[height=2in]{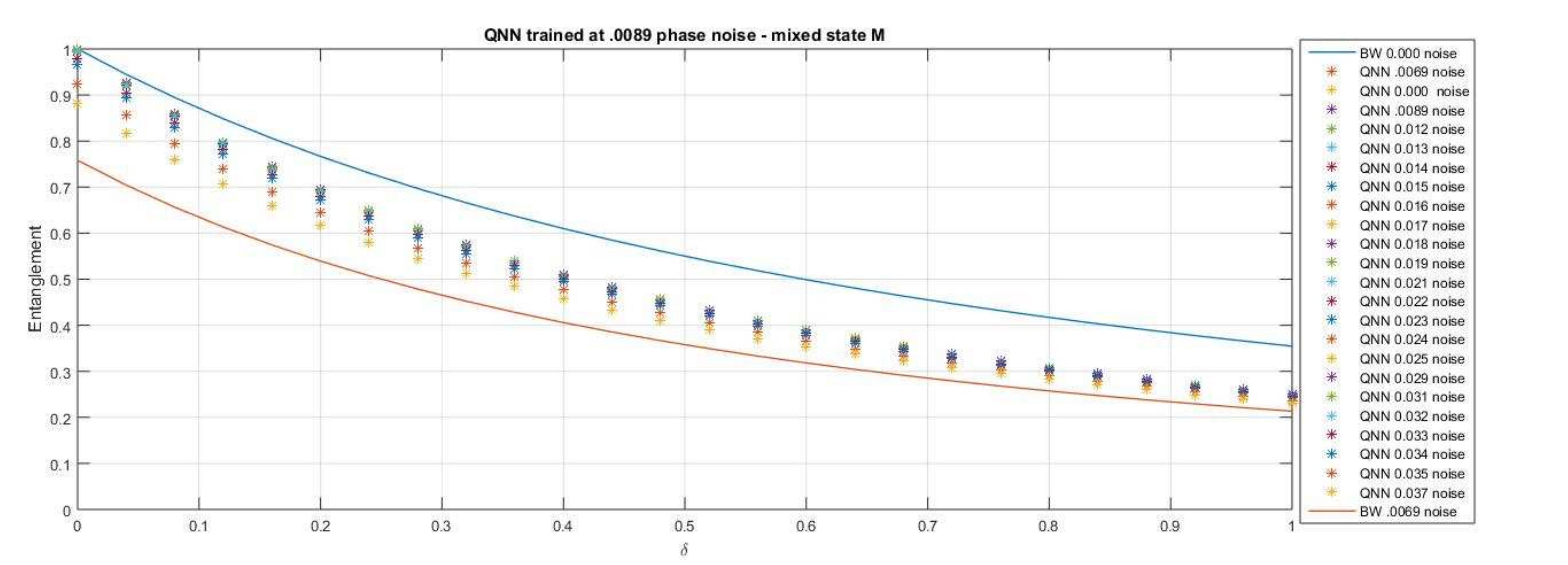} 
\caption{Entanglement of the state $M$ as a function of $\delta$, as calculated by the QNN, and compared with the entanglement of formation (marked ``BW'') at zero phase noise (blue) and at 0.0069 phase noise (orange). In each case the QNN was trained at a phase noise level of 0.0089, and then tested at the given level.}
\label{M89ph}
\end{figure}

\begin{figure}
\includegraphics[height=2in]{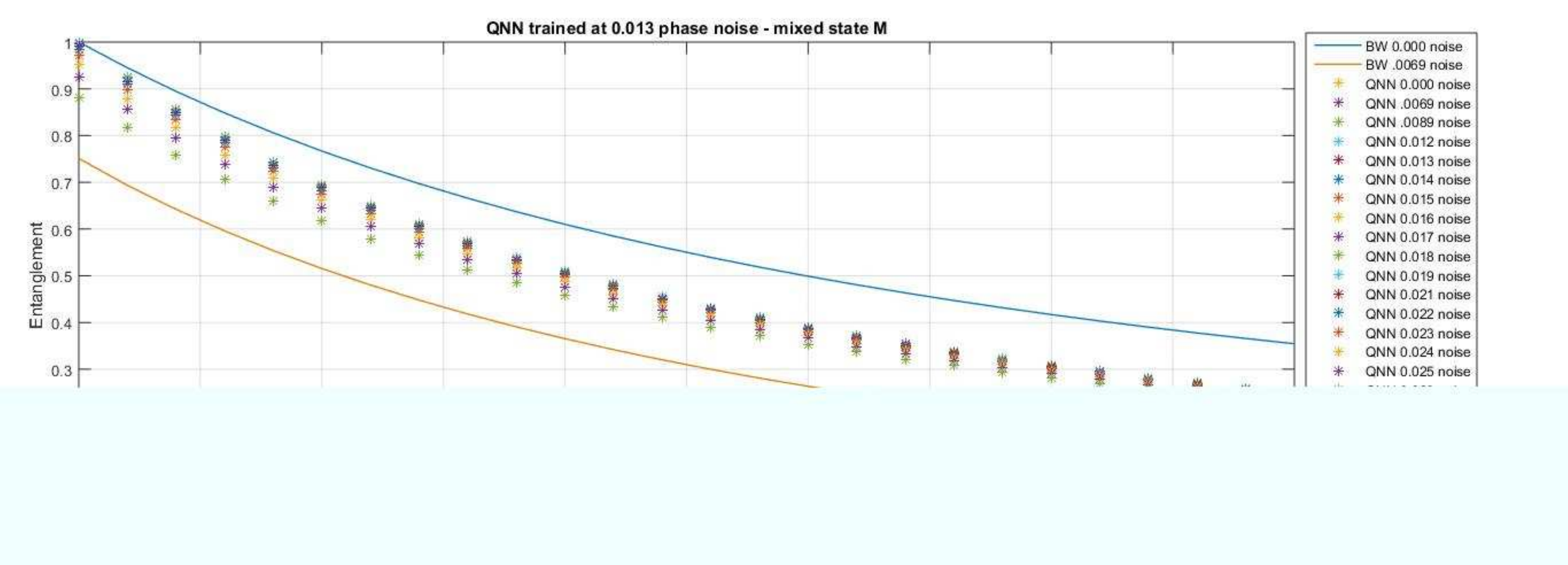} 
\caption{Entanglement of the state $M$ as a function of $\delta$, as calculated by the QNN, and compared with the entanglement of formation (marked ``BW'') at zero phase noise (blue) and at 0.0069 phase noise (orange). In each case the QNN was trained at 0.013 phase noise, and then tested at the given level.}
\label{M13ph}
\end{figure}

Figures~\ref{M0ph},\ref{M89ph},\ref{M13ph} show the performance of the QNN on the mixed state $M$, as trained, respectively, at zero, 0.0089, and 0.013 decoherence (phase noise), and tested with various levels of phase noise, and compared with the entanglement of formation at zero and at 0.0069 phase noise. Again, we see that the QNN entanglement indicator is robust to decoherence.


\section{ Noise plus decoherence} 

\begin{figure}
\includegraphics[height=2in]{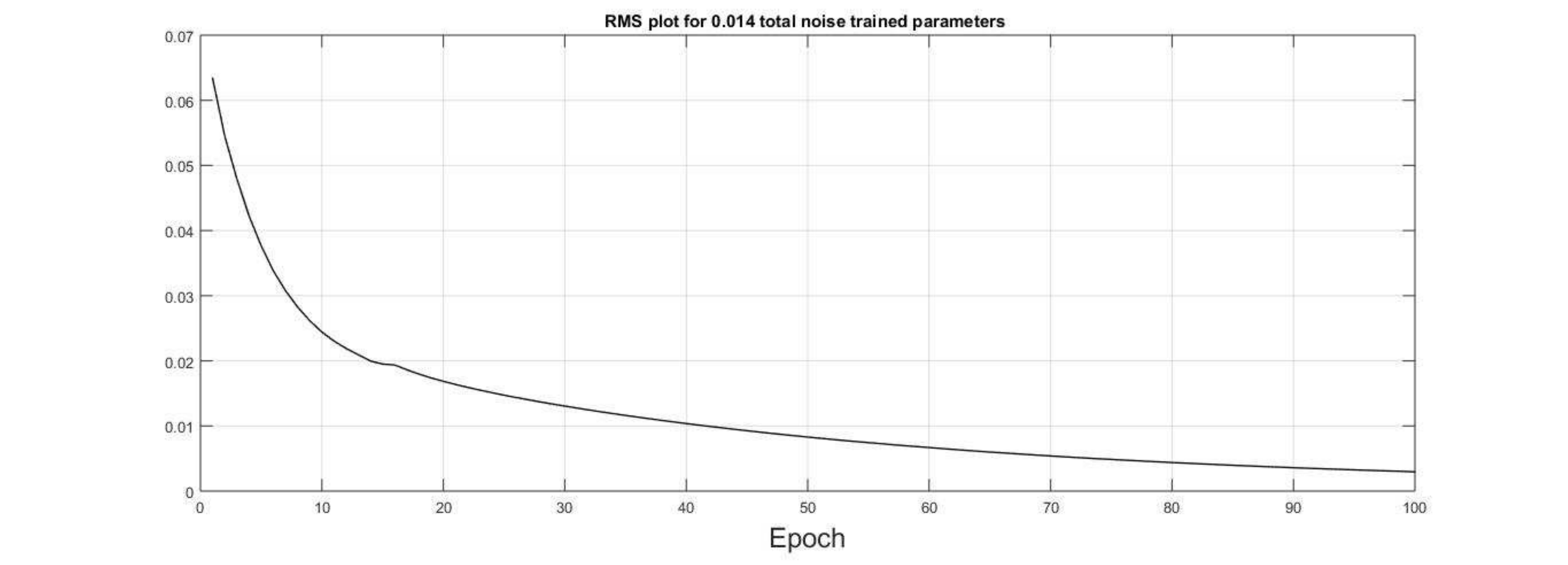} 
\caption{Total root mean squared error for the training set as a function of epoch (pass through the training set), for the 2-qubit system, with a complex noise level of 0.014 at each (of 317 total) timestep. Asymptotic error is $3.0 \times 10^{-3}$, approximately the same as with only magnitude noise.}
\label{cxtrn}
\end{figure}

\begin{figure}
\includegraphics[height=2in]{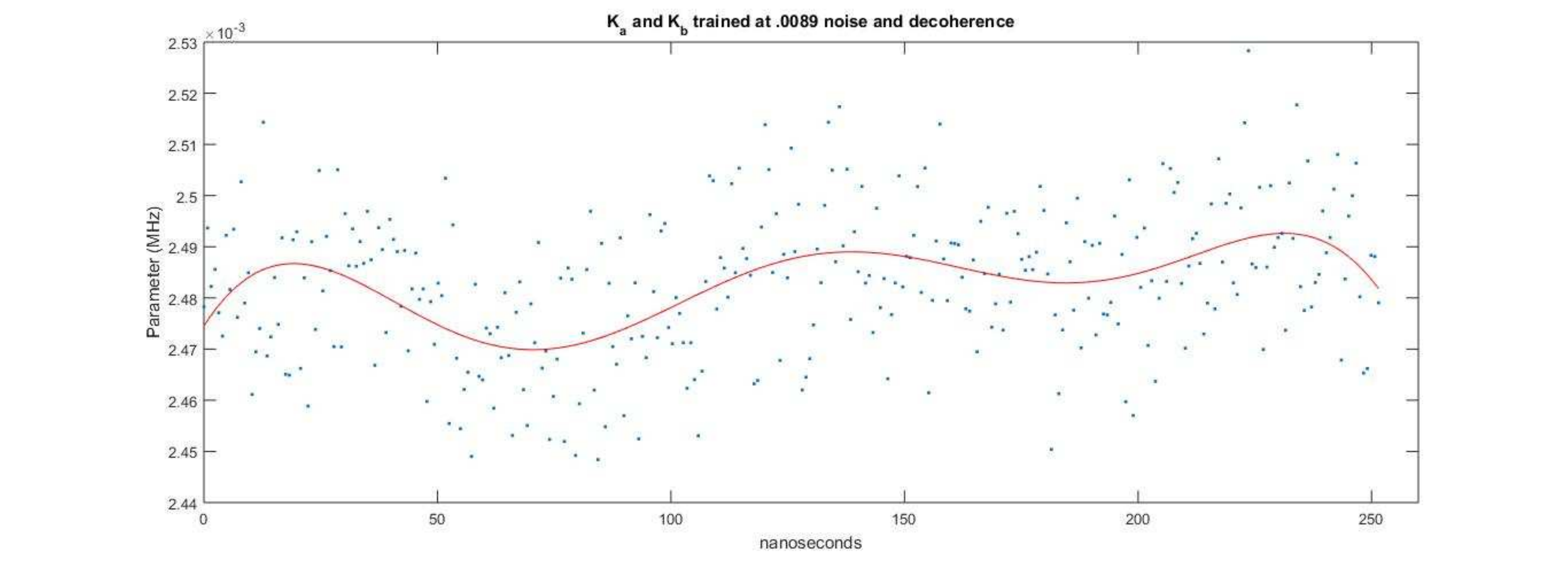} 
\caption{Parameter function $K_{A}=K_{B}$ as a function of time, as trained at 0.0089 complex noise at each of the 317 timesteps, for the entanglement indicator (data points), and plotted with the Fourier fit (solid line). Note the change in scale from Figure~\ref{K0}, because of the (much larger) spread of the noisy data: the Fourier fit is actually almost the same on this graph.}
\label{Kcx}
\end{figure}

\begin{figure}
\includegraphics[height=2in]{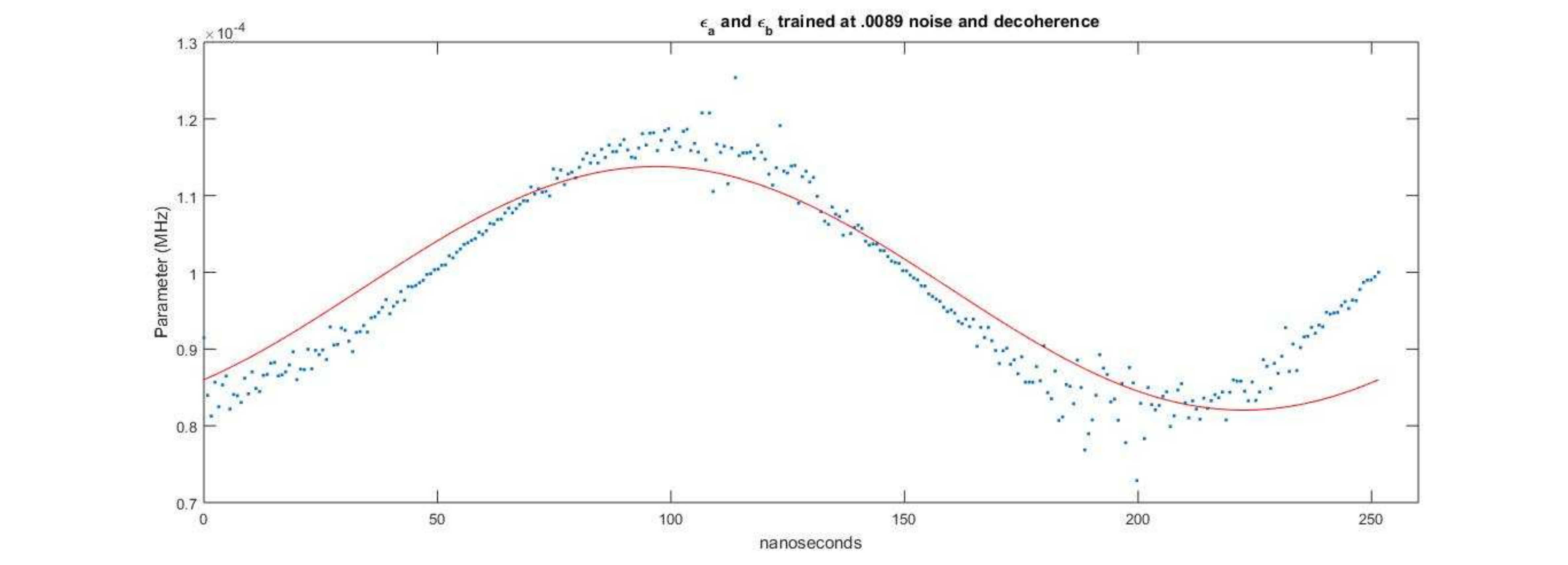} 
\caption{Parameter function $\epsilon_{A}=\epsilon_{B}$ as a function of time, as trained at 0.0089  complex noise at each of the 317 timesteps,  for the entanglement indicator (data points), and plotted with the Fourier fit (solid line).}
\label{epcx}
\end{figure}

\begin{figure}
\includegraphics[height=2in]{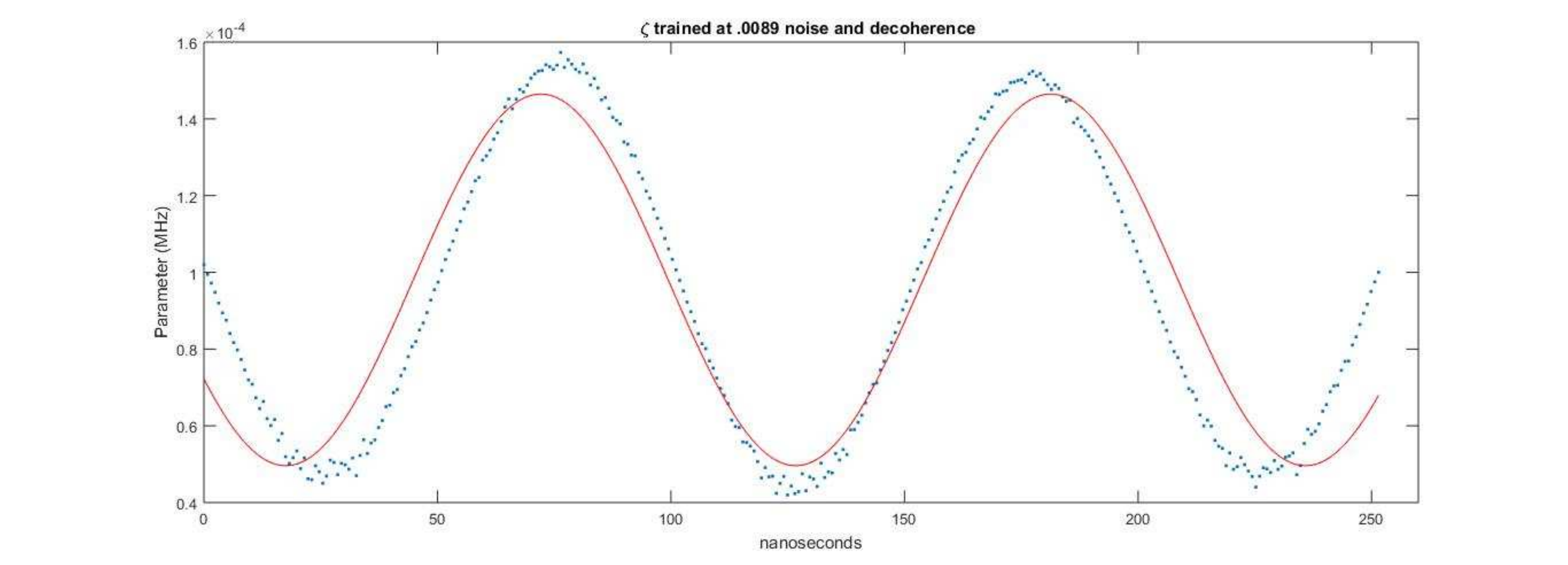} 
\caption{Parameter function $\zeta$ as a function of time, as trained at 0.0089 complex noise at each of the 317 timesteps,  for the entanglement indicator (data points), and plotted with the Fourier fit (solid line).}
\label{zcx}
\end{figure}

Finally, we consider the case of noise plus decoherence, that is, what we are calling random complex noise. For this case, we add both magnitude and phase noise. Figure~\ref{cxtrn} shows a typical rms error training curve for a the same level of complex noise as in Figures ~\ref{magtrn} and \ref{phtrn}. Again, asymptotic error for the training set does increase with increasing complex noise. The parameter functions again become ``noisy'': See Figures~\ref{Kcx},\ref{epcx},\ref{zcx}. 


\begin{figure}
\includegraphics[height=2in]{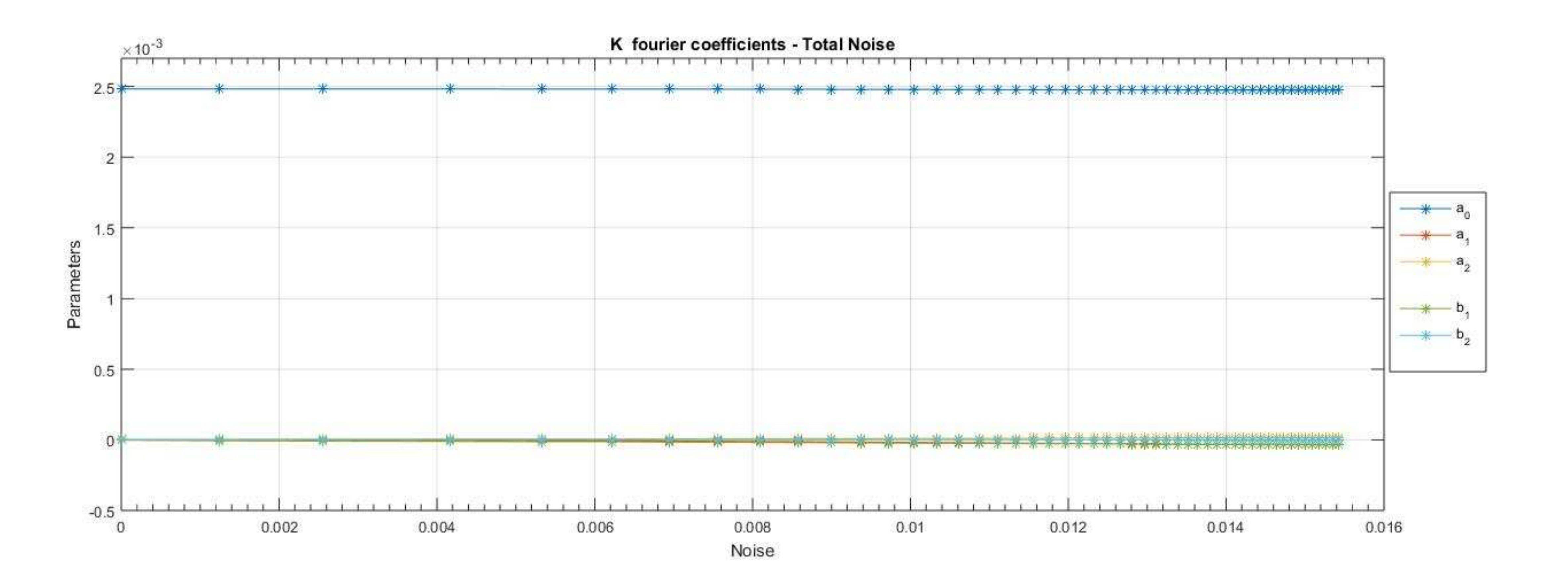} 
\caption{Fourier coefficients for the tunneling parameter functions $K$, as functions of complex noise level.}
\label{FKcx}
\end{figure}

\begin{figure}
\includegraphics[height=2in]{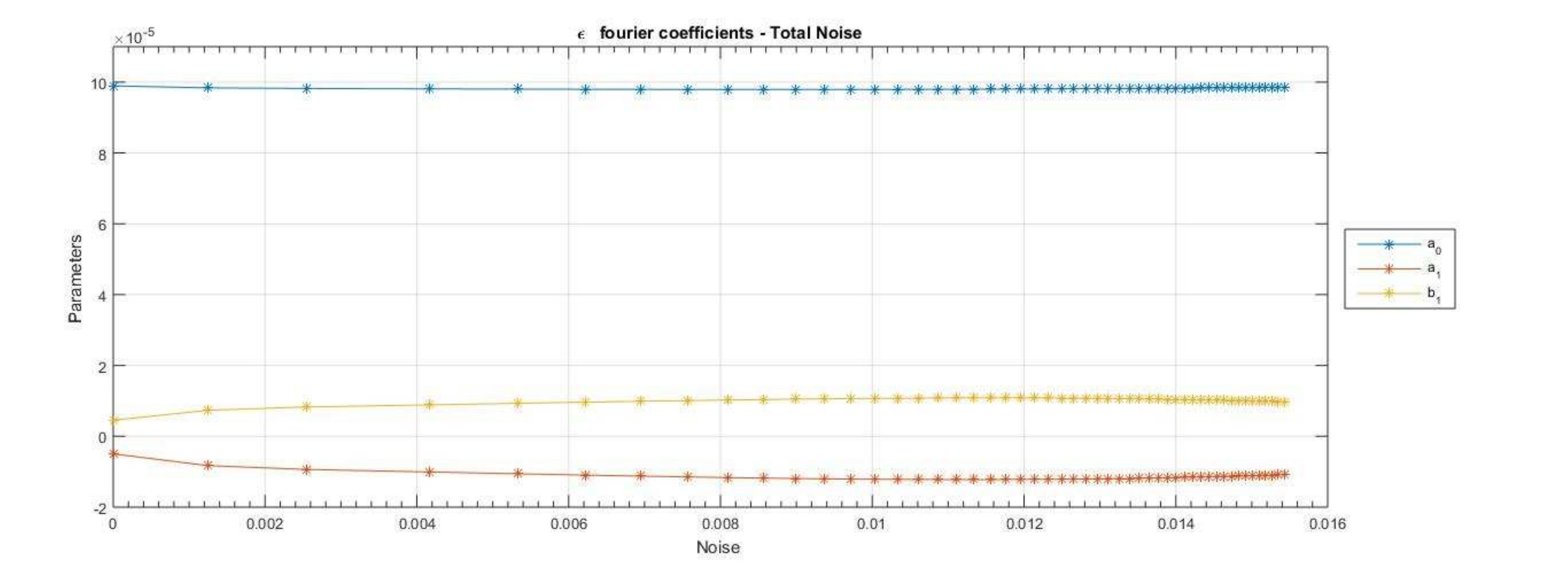} 
\caption{Fourier coefficients for the bias parameter functions $\epsilon$, as functions of complex noise level.}
\label{Fepcx}
\end{figure}

\begin{figure}
\includegraphics[height=2in]{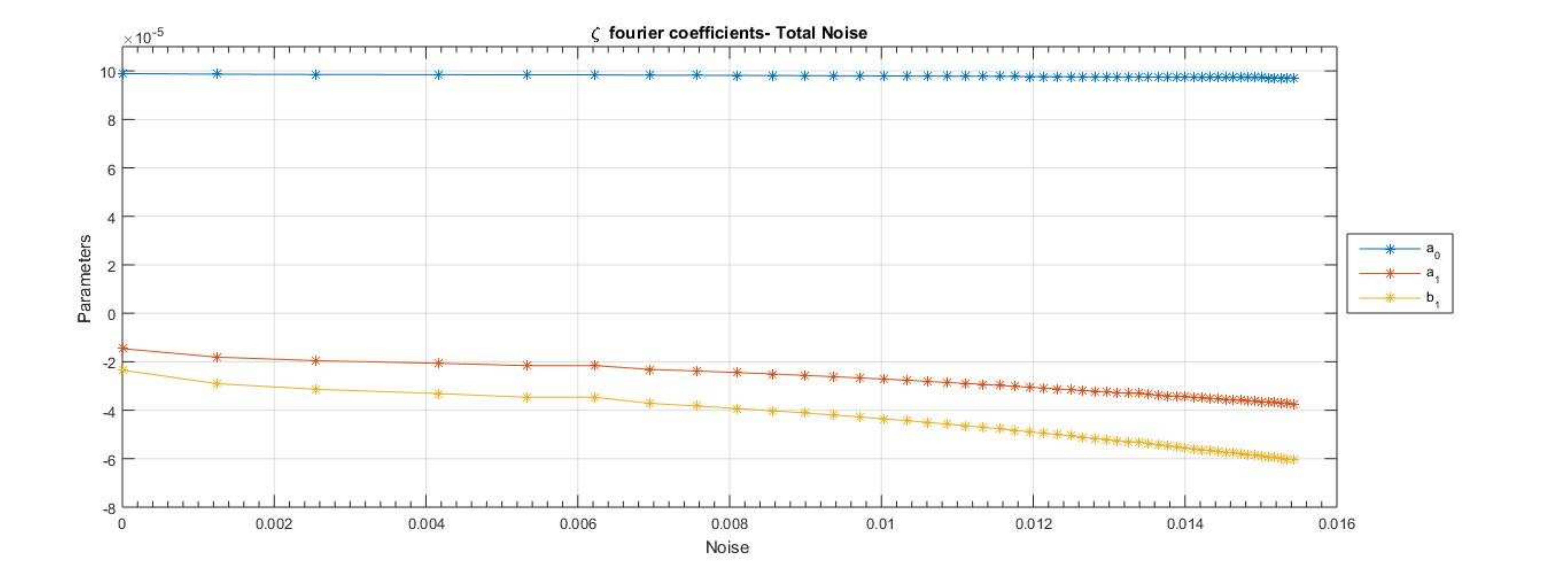} 
\caption{Fourier coefficients for the coupling parameter function $\zeta$, as functions of complex noise level.}
\label{Fzcx}
\end{figure}

Again, we test to see how much the Fourier fit changes, this time with complex noise. Figures~\ref{FKcx},\ref{Fepcx},\ref{Fzcx} show the Fourier coefficients as a function of complex noise level, for $K$, $\epsilon$, and $\zeta$, respectively. We can see that the indicator is, again, relatively stable.


\begin{figure}
\includegraphics[height=2in]{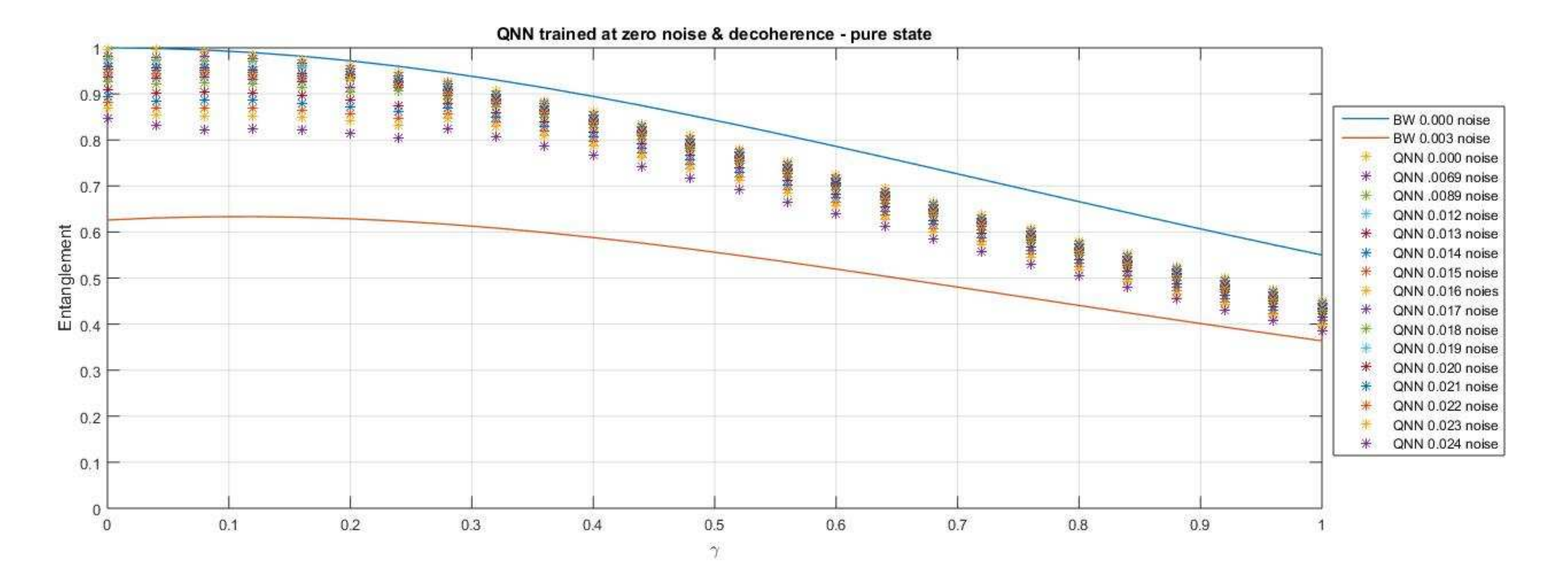} 
\caption{Entanglement of the state $P$ as a function of $\gamma$, as calculated by the QNN, and compared with the entanglement of formation (marked ``BW'') at zero noise (blue) and at 0.69\% noise plus decoherence (orange). In each case the QNN was trained at zero noise, but tested at the given level.}
\label{P0cx}
\end{figure}

\begin{figure}
\includegraphics[height=2in]{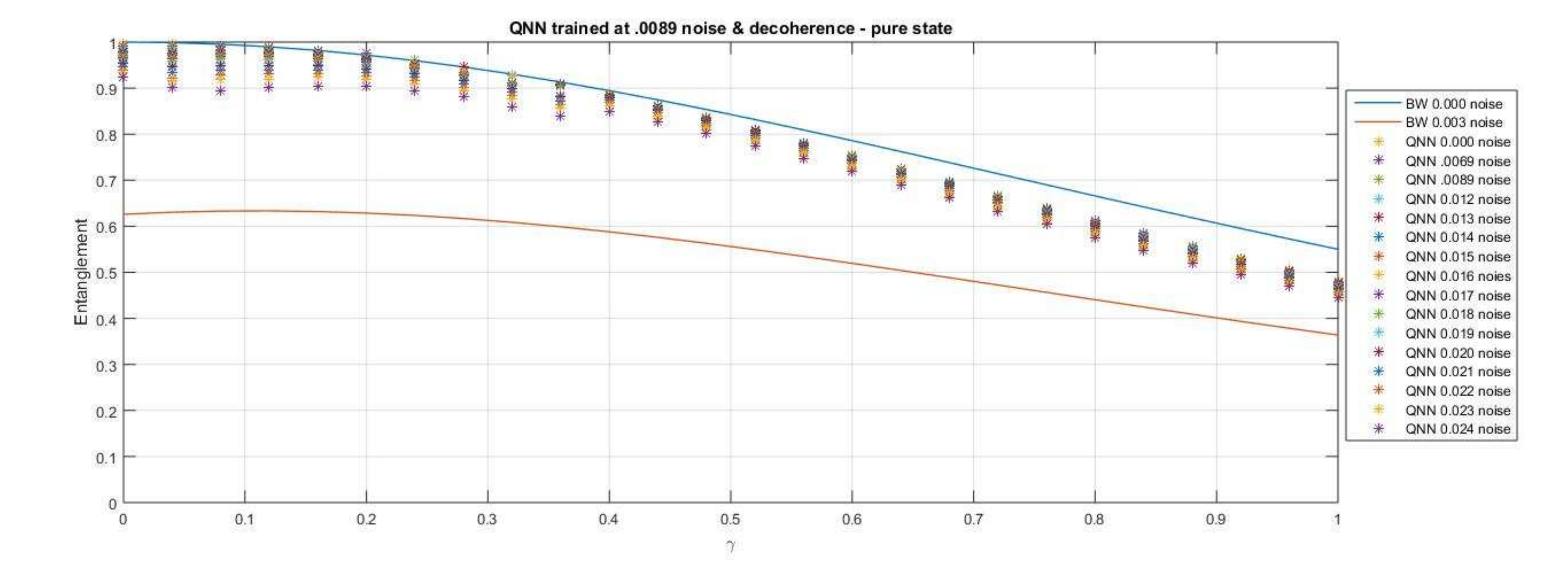} 
\caption{Entanglement of the state $P$ as a function of $\gamma$, as calculated by the QNN, and compared with the entanglement of formation (marked ``BW'') at zero noise (blue) and at 0.0069 noise plus decoherence (orange). In each case the QNN was trained at a noise level of 0.0089 complex noise, and then tested at the given level.}
\label{P89cx}
\end{figure}

\begin{figure}
\includegraphics[height=2in]{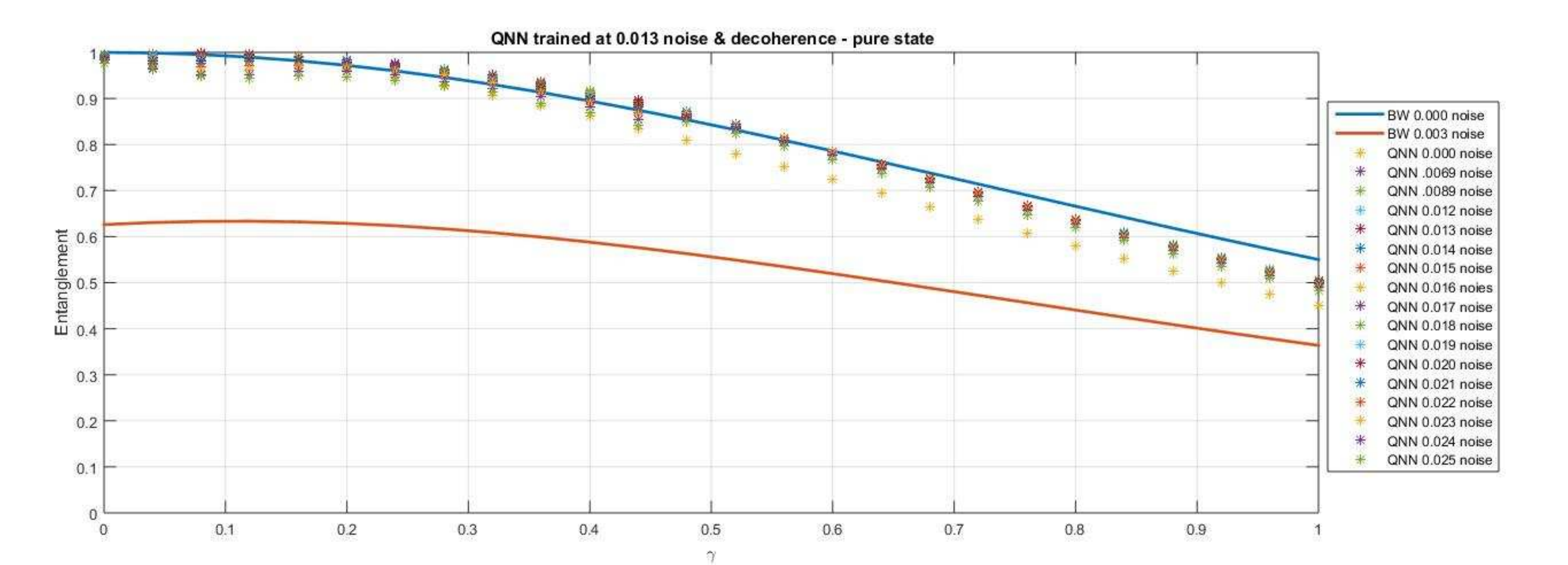} 
\caption{Entanglement of the state $P$ as a function of $\gamma$, as calculated by the QNN, and compared with the entanglement of formation (marked ``BW'') at zero complex noise (blue) and at 0.0069 noise plus decoherence(orange). In each case the QNN was trained at 0.013 level complex noise, and then tested at the given level.}
\label{P13cx}
\end{figure}

\begin{figure}
\includegraphics[height=2in]{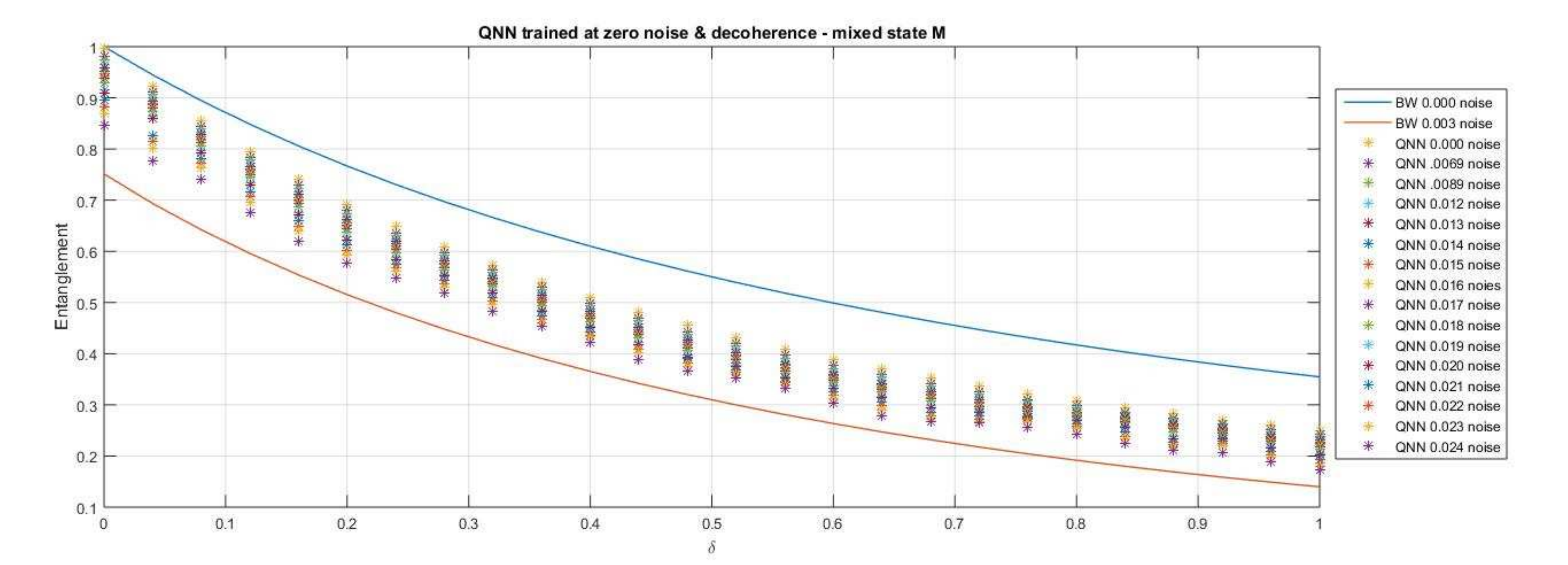} 
\caption{Entanglement of the state $M$ as a function of $\delta$, as calculated by the QNN, and compared with the entanglement of formation (marked ``BW'') at zero noise (blue) and at 0.0069 noise plus decoherence(orange). In each case the QNN was trained at zero noise, but tested at the given level.}
\label{M0cx}
\end{figure}

\begin{figure}
\includegraphics[height=2in]{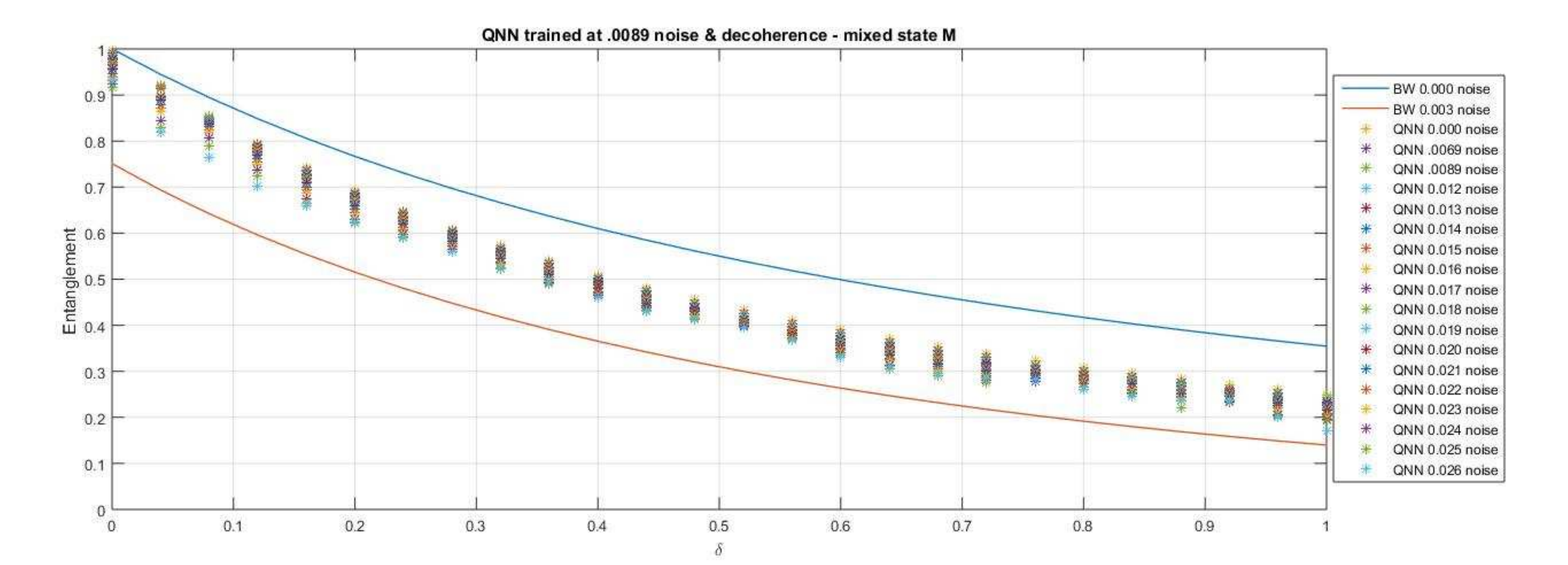} 
\caption{Entanglement of the state $M$ as a function of $\delta$, as calculated by the QNN, and compared with the entanglement of formation (marked ``BW'') at zero noise (blue) and at 0.0069 noise plus decoherence (orange). In each case the QNN was trained at a complex noise level of 0.0089, and then tested at the given level.}
\label{M89cx}
\end{figure}

\begin{figure}
\includegraphics[height=2in]{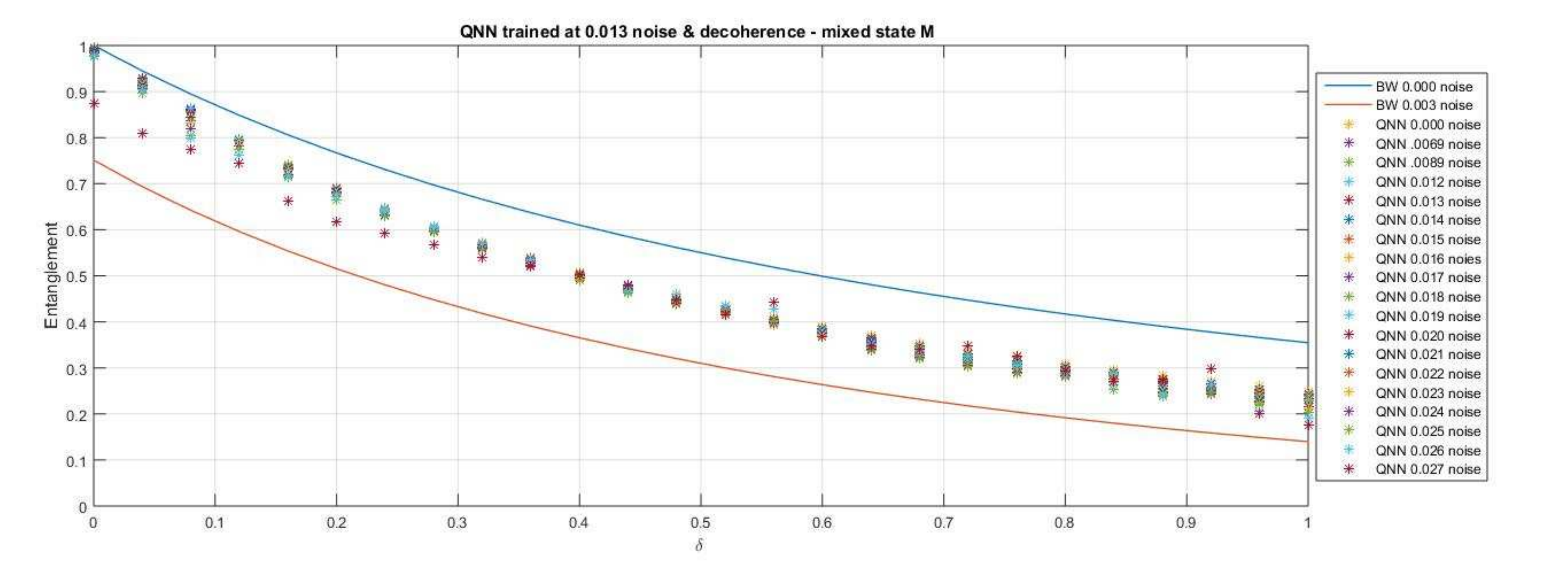} 
\caption{Entanglement of the state $M$ as a function of $\delta$, as calculated by the QNN, and compared with the entanglement of formation (marked ``BW'') at zero noise (blue) and at 0.0069 noise plus decoherence (orange). In each case the QNN was trained at 0.013 complex noise, and then tested at the given level.}
\label{M13cx}
\end{figure}

Again we use the Fourier fitted functions to test on both pure and mixed states. Figures~\ref{P0cx},\ref{P89cx},\ref{P13cx} show performance of the QNN on the entanglement of the state $P$, as trained, respectively, at zero, 0.0089, and 0.013 amplitude complex noise, and tested with various levels of complex noise, and compared with the entanglement of formation at zero and at 0.0069 complex noise. Figures~\ref{M0cx}~\ref{M89cx}~\ref{M13cx} show the performance of the QNN on the mixed state $M$, as trained, respectively, at zero, 0.0089, and 0.013 complex noise, and tested with various levels of complex noise, and compared with the entanglement of formation at zero and at 0.0069 complex noise. Results are excellent; in fact, they are somewhat better than for the case of only ``magnitude'' noise. In some sense allowing for decoherence makes the indicator even more robust. 



\section{ Conclusions} 

In previous work, we have proposed an entanglement indicator for a general qubit system. This indicator is a quantum system that processes the state whose entanglement is to be estimated. The parameters of the quantum system are adjusted via a supervised learning process using a sparse training set of states whose entanglement is well-defined. The learning is continued until sufficient training is achieved.  The trained parameter functions are well represented by single frequency functions (first order Fourier curve fit.)

We have shown here that those functions are robust to fairly high levels of noise and decoherence. The quantum neural network tests well on ``unknown'' states, both pure and mixed. Performance on decoherence in particular was excellent. We are reasonably confident that our results show that quantum neural networks are well suited for dealing with these types of problems in quantum computing. 

We are currently working to extend our results on noise and decoherence to multiple-qubit systems, using the well-known neural network technique of bootstrapping \cite{efron}. We also wish to understand exactly why the QNN is so robust to noise and decoherence. Classical neural networks are robust to noise and single neuron/synapse failure because of the multiple-redundance of parallel computing. Here, we have only a very small number of qubits/neurons, but we have designed our quantum network as operating over propagation in time, which can be written as a superposition of a very large number of definite time paths, using the Feynman path integral representation of quantum mechanics \cite{feynman}. In this picture, the instantaneous states of the quantum system at intermediate times, which are integrated over, play the role of ``virtual neurons''\cite{behrman}. In other words, it is possible that quantum superposition ensures redundance, even when the physical number of qubits is small, and thereby supplies fault-tolerance.  

\section{Acknowledgements}
\noindent
This work was supported in part by the National Science Foundation under Grant No. NSF PHY05-51164, through the KITP Scholars program (ECB), at the Kavli Institute for Theoretical Physics, University of California at Santa Barbara, Santa Barbara, CA.


\begin{thebibliography}{22}

\bibitem{bennett2}
 C.H. Bennett, D.P. DiVincenzo, J.A. Smolin, and W.K. Wootters (1996), 
{\it Mixed-state entanglement and quantum error correction}, Phys. Rev. A 54, pp. 3824-3851.

\bibitem{wootters}
 W.K. Wootters (1998), {\it Entanglement of formation of an arbitrary state of two qubits}, Phys. Rev. Lett. 80, pp. 2245-2248.
 
\bibitem{ghz}
 D.M. Greenberger, M.A. Horne, and A. Zeilinger (1989), in {\it Bell's Theorem and the Conception of the Universe}, M. Kafatos, ed. , Kluwer Acdemic (Dordrecht), p 107.

\bibitem{vedral}
 V. Vedral, M.B. Plenio, M.A. Rippin, and P.L. Knight (1997), {\it Quantifying entanglement}, 
Phys. Rev. Lett. 78, pp. 2275-2279; V Vedral and M.B. Plenio (1998),
{\it Entanglement measures and purification procedures}, Phys. Rev. A 57, pp. 1619-1633; 
L. Henderson and V. Vedral (2001), {\it Classical, quantum and total correlations}, J. Phys. A 34, pp.  6899-6905.

\bibitem{tamaryan} 
S. Tamaryan, A. Sudbery, and L. Tamaryan (2010), {\it Duality and the geometric measure of entanglement of general multiqubit W states}, Phys. Rev. A 81, 052319.

\bibitem{park}
H.S. Park, S.-S.B. Lee, H. Kim, S.-K. Choi, and H.-S. Sim (2010), {\it Construction of optimal witness for unknown two-qubit entanglement}, Phys. Rev. Lett. 105, 230404.

\bibitem{filip}
 R. Filip (2002), {\it  Overlap and entanglement-witness measurements }, Phys. Rev. A 65, 062320; F.G.S.L. Brando (2005), {\it Quantifying entanglement with witness operators }, quant-ph/0503152.
 
\bibitem{yamamoto}
 T. Yamamoto, Yu.A. Pashkin, O. Astafiev, Y. Nakamura, and J.S. Tsai (2003), 
{\it Demonstration of conditional gate operation using superconducting charge qubits},  Nature 425, pp. 941-944.

\bibitem{lecun}
Yann le Cun (1988), {\it A theoretical framework for back-propagation} in 
{\it Proc. 1998 Connectionist Models Summer School,} D. Touretzky, G. Hinton, and T. Sejnowski, eds., Morgan Kaufmann, (San Mateo), pp. 21-28.

\bibitem{werbos}
Paul Werbos (1992), in {\it Handbook of Intelligent Control }, Van Nostrand Reinhold, pp. 79-80 and 339-344.

\bibitem{behrmanqic}
E.C. Behrman, J.E. Steck, P. Kumar, and K.A. Walsh (2008), {\it Quantum algorithm design using dynamic learning}, Quantum Information and Computation 8, pp. 12-29.

\bibitem{efron}
B. Efron and R.J. Tibshirani (1994), {\it An Introduction to the bootstrap}. Boca Raton, FL: Chapman and Hall/CRC.

\bibitem{nabic}
E.C. Behrman and J.E. Steck, {\it Dynamic learning of pairwise and three-way entanglement}, in {\it Proceedings of the Third World Congress on Nature and Biologically Inspired Computing (NaBIC 2011)} (Salamanca, Spain, October 19-21, 2011.) (Institute of Electrical and Electronics Engineers).

\bibitem{behrmanmulti}
E.C. Behrman and J.E. Steck, {\it Multiqubit entanglement of a general input state}, Quantum Information and Computation {\bf 13}, 36-53 (2013).

\bibitem{wasserman}
P.D. Wasserman (1993), {\it Advanced Methods in Neural Computing}. New York: Van Nostrand Reinhold.

\bibitem{singapore}
E.C. Behrman and J.E. Steck, {\it A quantum neural network computes its own relative phase}, in {\it Proceedings of the IEEE Symposium on Computational Intelligence 2013} (Singapore, April 15-19, 2013.) (Institute of Electrical and Electronics Engineers).

\bibitem{behrmanieee}
E.C. Behrman, R.E.F. Bonde, J.E. Steck, and J.F. Behrman (2014), {\it On the correction of anomalous phase oscillation in entanglement witnesses using quantum neural networks}, IEEE Transactions on Neural Networks and Learning Systems 25 (9), pp.1696-1703.

\bibitem{peres}
A. Peres (1995), {\it Quantum Theory: Concepts and Methods}. Dordrecht, The Netherlands: Kluwer.

\bibitem{feynman}
R.P. Feynman (1951), {\it An operator calculus having applications in quantum electrodynamics}, Phys. Rev. 84, 108-128; 
R.P. Feynman and A.R. Hibbs (1965), {\it Quantum Mechanics and Path Integrals}, McGraw-Hill (New York).

\bibitem{behrman}
E. C. Behrman, L. R. Nash, J. E. Steck, V. G. Chandrashekar, and S. R. Skinner (2000), 
{\it Simulations of quantum neural networks}, Information  Sciences 128,  pp. 257-269.

\end{thebibliography}
\end{document}